\documentclass[axodraw]{JHEP3}

\usepackage{graphicx}
\usepackage{graphics}
\usepackage{epsfig}
\epsfclipon

\usepackage{epsfig,bm}
\usepackage{latexsym}
\usepackage{amssymb,amsmath}

\usepackage{axodraw}

\usepackage[small]{caption}

\usepackage{amsfonts}

\newcommand{\be}{\begin{equation}}
\newcommand{\ee}{\end{equation}}
\newcommand{\bea}{\begin{eqnarray}}
\newcommand{\eea}{\end{eqnarray}}

\def\lsim{\mathrel{\rlap{\lower4pt\hbox{\hskip1pt$\sim$}}
    \raise1pt\hbox{$<$}}}                
\def\gsim{\mathrel{\rlap{\lower4pt\hbox{\hskip1pt$\sim$}}
    \raise1pt\hbox{$>$}}}                

\def\ie{{\it i.e.\,}}

\def\revise#1       {\raisebox{-0em}{\rule{3pt}{1em}}%
                     \marginpar{\raisebox{.5em}{\vrule width3pt\
                     \vrule width0pt height 0pt depth0.5em
                     \hbox to 0cm{\hspace{0cm}{%
                     \parbox[t]{4em}{\raggedright\footnotesize{#1}}}\hss}}}}

\def\del          {\partial}

\def\sqr#1#2{{\vcenter{\vbox{\hrule height.#2pt
 \hbox{\vrule width.#2pt height#1pt \kern#1pt
 \vrule width.#2pt}\hrule height.#2pt}}}}

\def\aa1{\phi}
\def\cc1{\psi}

\newcommand{\eq}{\begin{equation}}
\newcommand{\eqx}{\end{equation}}
\newcommand{\eqn}{\begin{eqnarray}}
\newcommand{\eqnx}{\end{eqnarray}}

\newcommand{\ud}{\,\mathrm{d}}

\catcode`\@=12

\title{Reheating the Universe After Multi-Field Inflation}

\author{Jonathan Braden$ ^{a,b,1}$, Lev Kofman$ ^{a}$ and Neil Barnaby$ ^{a,2}$\\
\it $ ^a$Canadian Institute for Theoretical Astrophysics, University of Toronto,\\
\it  60 St. George St., Toronto, ON M5S 3H8, Canada\\
\it $ ^b$Department of Physics, University of Toronto,\\
\it 60 St. George St., Toronto, ON M5S 1A7, Canada \\
$^{1}$jbraden@physics.utoronto.ca $^{2}$barnaby@cita.utoronto.ca
}

\abstract{ 
We study in detail (p)reheating after multi-field inflation models with a particular
focus on N-flation.  We consider a variety of different couplings between the inflatons
and the matter sector, including both quartic and trilinear interactions with a light
scalar field.  We show that the presence of multiple oscillating inflatons makes parametric
resonance inefficient in the case of the quartic interactions.  Moreover, perturbative processes
do not permit a complete decay of the inflaton for this coupling.  In order to recover the hot big
bang, we must instead consider trilinear couplings.  In this case we show that strong nonperturbative
preheating is possible via multi-field tachyonic resonance.  In addition, late-time perturbative effects do permit
a complete decay of the condensate. We also study the production of gauge fields for several prototype 
couplings, finding similar results to the trilinear scalar coupling.  During the course of our analysis
we develop the mathematical theory of the quasi-periodic Mathieu equation, the multi-field generalization
of the Floquet theory familiar from preheating after single field inflation.  We also elaborate on the theory of
perturbative decays of a classical inflaton condensate, which is applicable in single-field models also.
}

\preprint{}

\keywords{string theory inflation, (p)reheating, N-flation, assisted inflation}

\begin{document}

\section{Introduction}
\label{sec:intro}

Reheating after inflation is a crucial requirement for any model.  Depending on how inflation ends
and how the inflaton interacts, the transfer of energy from the inflaton sector into standard model (SM) degrees of freedom
may proceed differently.  The detailed physics of (p)reheating is rather model dependent and previous studies have uncovered
a wide array of different quantum field theory (QFT) phenomena, including perturbative \cite{kofman25plus} and nonperturbative \cite{KLS94,KLS97,frolov} production of both bosons 
and fermions \cite{ferm}.  At the nonperturbative level a rich variety of preheating processes are possible, including parametric resonance \cite{KLS94,KLS97,res}, tachyonic
preheating \cite{tac} and tachyonic resonance \cite{tac_res}.  Examples have also been studied where the dynamics of reheating involves distinctively string theoretic
processes \cite{BBC,KY,BBHK} and cannot be captured by ordinary QFT.  (See also \cite{stringy_reheating} for further studies of string theory 
reheating and \cite{post_inf} for other aspects of the endpoint of string theory inflation.)

The requirement of successful reheating places a number of nontrivial constraints on inflationary model building.  At the most basic level, one must ensure
that the inflaton field completely decays into radiation after inflation, in order to match smoothly onto the usual hot big bang evolution.  Moreover, one
must verify that exotic relics (which might interfere with the successful predictions of big bang nucleosynthesis) are not over-produced.  It is therefore
crucial to understand the decay channels of the inflaton in order to assess the viability of any inflation model.  Depending on the model, the process of preheating 
may also lead to various cosmological observables such as gravitational waves \cite{gw} or nongaussianities \cite{ng1,ng2}.

Previous studies of (p)reheating after inflation have focused largely on the decay of a small number of fields in the inflaton sector \cite{KLS97}.  However, there are strong motivations
to consider inflation models in which a huge number of fields play an important dynamical role.  One might expect such models to arise naturally in string theory,
where realistic compactifications often contain exponentially large numbers of moduli and axion fields \cite{bigN,Nflation,matrix_spectrum}.  Indeed, 
string theory appears to contain a rich landscape \cite{Susskind} and, in general, stringy inflation might involve the motion of a huge 
number of fields on a very complicated scalar potential \cite{multi_land,staggered,rand1,rand2,stream}.  Only in very special corners of this cosmic landscape
would one expect inflation to take place along a single direction in field space.  

Multi-field inflation is also desirable from a phenomenological perspective.  The ``assisted inflation'' mechanism (first discovered in \cite{assisted} and generalized to a much broader 
class of models in \cite{assisted2}) allows the collective motion of a large number of scalar fields $\phi_i$ ($i = 1,\cdots, N$) to
drive inflation, even when each individual field could not support inflation on its own.  The essence of 
this mechanism is very simple. Suppose the fields are decoupled
\begin{equation}
\label{sumV}
  V(\phi_j) = \sum_{i=1}^N V_i(\phi_i) \, .
\end{equation}
Then each inflaton feels the Hubble friction induced by the collective potential energies
\begin{equation}
\label{fried}
  3 H^2 = \frac{1}{M_p^2} \sum_{i=1}^N \left[ V_i(\phi_i) + \frac{\dot{\phi}_i^2}{2} \right] \cong \frac{1}{M_p^2} \sum_{i=1}^N  V_i(\phi_i)
\end{equation}
but only the restoring force coming from its own potential
\begin{equation}
\label{KG}
  \ddot{\phi}_i + 3 H \dot{\phi}_i = - V'_i(\phi_i) \, .
\end{equation}
With sufficiently large $N$ the motion of each $\phi_i$ will be over-damped, even if each $V_i(\phi_i)$ is relatively steep, alleviating (to some extent)
the fine-tuning problems associated with obtaining flat scalar potentials (see, for example, \cite{flat_scalar_pot}).
This mechanism permits a realization of chaotic inflation in a regime where each $\phi_i$ undergoes a sub-Planckian displacement 
in field space, whereas the collective radial excitation $\rho$
(defined by $\rho^2 = \sum_i \phi_i^2$) traverses a super-Planckian distance.  Such a model goes a long way to ease concerns 
about the validity of the effective field theory description of large field inflation \cite{lyth_bound}.

In \cite{Nflation} a particularly appealing realization of assisted inflation was proposed: \emph{N-flation}.  In this model one considers the dynamics of $N$ axion fields $\phi_i$, each associated with
a different broken shift symmetry $\phi_i \rightarrow \phi_i + c_i$.  In the region relevant for inflation, the potential takes the simple form (\ref{sumV}) 
with $V_i(\phi_i) \cong \frac{m_i^2}{2}\phi_i^2$.  The potential of N-flation has been argued to be radiatively stable \cite{Nflation}.  (See also \cite{Nflation2,Nflation3,Nflation4} for more details on the dynamics/phenomenology of this model.)  
N-flation is quite natural from the string theory perspective.  In oriented critical string theories the massless 2-form field $B_{\mu\nu}$ can wrap a huge number of independent 2-cycles when the theory is 
compactified to four dimensions.  Each such cycle leads to an axion at low energies.  Explicit examples are known where the number of low energy axions may be as large as $N = \mathcal{O}(10^5)$ \cite{bigN}.  
Using random matrix theory, it has been shown that the predictions of N-flation are largely independent of the details of the compactification \cite{matrix_spectrum}.  See, however, \cite{KSS} for complications associated with consistently
embedding this framework into stabilized string theory vacua.

In this paper we study (p)reheating after multi-field inflation.  (See also \cite{multi_preheat,multi_preheat2}.)  We focus on N-flation as a specific example, however, many of our results will apply to multi-field models more generally.
We proceed phenomenologically, considering a variety of different couplings between the inflatons and matter.  For simplicity, we first consider reheating into a scalar field $\chi$ through prototype interactions of the form
\begin{equation}
  \mathcal{L}^{4-\mathrm{leg}}_{\mathrm{int}} = -\sum_{i=1}^N \frac{g_i^2}{2} \phi_i^2 \chi^2, \hspace{5mm} \mathcal{L}^{3-\mathrm{leg}}_{\mathrm{int}} = -\sum_{i=1}^N \frac{\sigma_i}{2} \phi_i \chi^2 \, .
\end{equation}
These two kinds of couplings yield very different results for the decay of the inflaton, both in the perturbative and nonperturbative regimes.  In the case of 4-leg interactions, nonperturbative preheating processes 
are suppressed by the de-phasing of the N-flaton oscillations.  We show that reheating must therefore proceed via perturbative decays for this choice of coupling.  This is in contrast to the single-field case, were nonperturbative
parametric resonance may occur.  On the other hand, we show that the 3-leg interactions \emph{do} allow for violent nonperturbative particle production via tachyonic resonance.  The presence of a huge number of oscillating
fields in the inflationary sector does \emph{not} prohibit preheating via such couplings.  During the course of our analysis we develop the mathematical theory of the quasi-periodic Mathieu equation, the generalization of the
usual Floquet theory to multi-field inflationary models.  We study quantitatively the structure of the stability/instability charts for multi-field inflation and comment on the dissolution of the stability bands in the presence
of multiple incommensurate frequencies.

Having ruled out nonperturbative preheating effects in the case of the 4-leg interactions, we next study the perturbative QFT decays of the coherent oscillating N-flatons.  We find that the 4-leg interaction does \emph{not} permit
a complete decay; some relic density of inflaton particles always freezes out and the late-time universe is typically cold and devoid of life as we know it.  This result is completely analogous to the single-field case, which we 
re-consider.

Next, we study the perturbative decays of the coherent inflaton oscillations via 3-leg interactions, which will be relevant during the final stages of reheating and also when the inflaton oscillations have small amplitude.  Here we show
that the inflaton condensate \emph{is} permitted to decay completely and the late-time universe may be compatible with the usual hot big bang scenario.  Again, this result is completely analogous to the single-field case.

Finally, we also consider coupling the N-flaton particles to gauge fields.  We consider the following prototype interactions between the inflatons $\phi_i$ 
and a $U(1)$ gauge field $A_\mu$ with field strength $F_{\mu\nu} = \partial_\mu A_\nu - \partial_\nu A_\mu$:
\begin{equation}
  \mathcal{L}^{\mathrm{axion}}_{\mathrm{int}} = -\sum_{i=1}^N\frac{\phi_i}{4 M_i} F^{\mu\nu} \tilde{F}_{\mu\nu}, \hspace{5mm} \mathcal{L}^{\mathrm{moduli}}_{\mathrm{int}} = -\sum_{i=1}^N\frac{\phi_i}{4 M_i} F^{\mu\nu} F_{\mu\nu}  \, .
\end{equation}
Here $\tilde{F}_{\mu\nu} = \frac{1}{2} \epsilon^{\mu\nu\rho\sigma} F_{\rho\sigma}$ is the dual field strength tensor, $\epsilon^{\mu\nu\rho\sigma}$ is the anti-symmetric Levi-Civita symbol and the $M_i$'s are symmetry breaking scales.
The former axion-type coupling to $F\wedge F$ is directly relevant for N-flation.  In this case we find that the production of gauge fields is very similar to the 3-legs scalar interaction above;
the inflaton is permitted to decay completely and nonperturbative preheating is possible via tachyonic resonance, depending on the model parameters $M_i$.  For completeness, we also consider the coupling $\sum_i\phi_i F^2/M_i$, which might be relevant
for string theory models in which inflation is realized via the motion of a large number of moduli fields. 

The organization of this paper is as follows. In section \ref{sec:Nflation} we review the N-flation model and its string theoretic realizations.  In section \ref{sec:preheating} we discuss the possibility of nonperturbative preheating
after N-flation.  During the course of our analysis, we develop the mathematical theory of the quasi-periodic Mathieu equation, a multi-field generalization of the Floquet analysis that is familiar in the preheating literature.
Next, we turn our attention to perturbative decays.  In section \ref{sec:pert} we review the perturbative decays of the coherent oscillations of the inflaton in single
field inflation and compare this to the more familiar case of decaying inflaton \emph{particles}.  In section \ref{sec:4legs} we discuss the perturbative decay of the N-flatons via 4-leg interactions, showing that the inflaton is not permitted 
to decay completely.  In section \ref{sec:3legs} we consider instead perturbative 3-leg interactions showing that, this time, a complete decay is possible.  In section \ref{sec:SM} we study the production of standard model gauge fields via both
perturbative and nonperturbative processes after N-flation.  In section \ref{sec:generalizations} we discuss some possible applications of our results outside of the context of N-flation and also study the production of gauge fields
via ``modular'' couplings to the inflatons.  Finally, in section \ref{sec:conclusions} we conclude and speculate on directions for future investigation.

\section{The N-flation Model}
\label{sec:Nflation}

\subsection{Homogeneous Dynamics}

In this section we briefly review the N-flation model and its stringy realization.  The idea of \cite{Nflation} was to realize multi-field assisted inflaton using axions.  
To first approximation, the potential takes the separable form (\ref{sumV}) where the $V_i$ are periodic and arise solely from nonperturbative effects:
\begin{equation}
  V_i(\phi_i) = \Lambda_i^4 \left[ 1 -  \cos\left(\frac{2\pi \phi_i}{f_i}\right)  \right] + \cdots \,\, .
\end{equation}
Here $f_i$ is the axion decay constant and $\Lambda_i$ is some dynamically generated scale.  For $\Lambda_i$ well below the ultra-violet (UV) scale 
we can neglect higher harmonics in the potential and also cross terms coming from multi-instanton effects, which might spoil the separable form (\ref{sumV}).
For small field values $\phi_i \ll f_i$ we can expand the potential $V_i$ as
\begin{equation}
\label{V_i}
  V_i(\phi_i) \cong \frac{m_i^2}{2}\phi_i^2 \, ,
\end{equation}
where $m_i = 2\pi \Lambda_i^2 / f_i$.  This potential provides a realization of chaotic inflation for the collective variable $\rho = \sqrt{\sum_i \phi_i^2}$, which may experience a super-Planckian displacement
in field space, even when the individual field displacements are sub-Planckian.  

\begin{figure}[h!]
    \centering
    \includegraphics[width=0.5\textwidth]{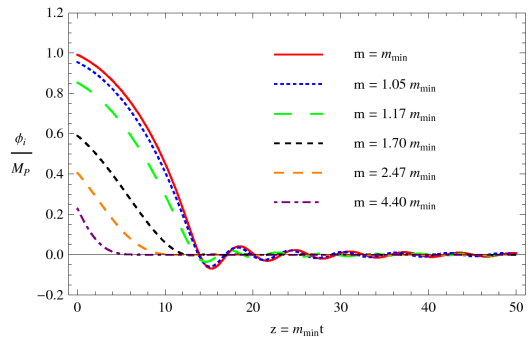}
\caption{Evolution of several N-flaton fields showing both the inflationary period and post-inflation oscillations. This figure illustrates the damping of heavier fields during inflation.  
At the onset of the oscillatory phase relevant for (p)reheating, only the lightest fields remain dynamical.  Notice that these are nearly degenerate in mass, and therefore their post-inflationary
oscillations are nearly in phase.  For illustration we take $N=1500$ fields with equal energy initial conditions at the beginning of inflation.\label{fig:field1500}}
\end{figure}

The homogeneous dynamics of the N-flaton fields are described by the usual Klein-Gordon equation
\begin{equation}
\label{NKG}
  \ddot{\phi}_i(t) + 3 H \dot{\phi_i}(t) + m_i^2 \phi_i(t) = 0 \, .
\end{equation}
During the slow roll phase the friction term dominates and one can easily see \cite{matrix_spectrum}
\begin{equation}
\label{spread}
  \frac{\phi_i(t)}{\phi_i(t_0)} = \left( \frac{\phi_j(t)}{\phi_j(t_0)} \right)^{m_i^2 / m_j^2} \, .
\end{equation}
This shows that even if the fields begin near the same point in field space they quickly spread out during inflation: the heaviest fields damp rapidly to the minimum and drop out of the inflationary dynamics
while the lightest fields remain frozen until the end of inflation.  We illustrate these dynamics in Fig.~\ref{fig:field1500}.

Once inflation has ended, the lightest N-flaton fields (those which have not been damped to zero during inflation) will oscillate about the global minimum of the potential $V = \sum_i \frac{m_i^2}{2}\phi_i^2$.  
It is this phase of post-inflationary oscillations which is relevant for (p)reheating.  To describe these oscillations, note that equation (\ref{NKG}) can be written as
\begin{equation}
  \frac{d^2}{dt^2}(a^{3/2} \phi_i) + \left[ m_i^2 - \left(\frac{9}{4} H^2 + \frac{3}{2}\dot{H}  \right)\right] (a^{3/2}\phi_i) = 0 \, .
\end{equation}
Since $m^2_i \gg H^2, \dot{H}$ during preheating the surviving N-flatons will oscillate as
\begin{equation}
\label{Nosc}
  \phi_i(t) \cong \frac{\phi_{0,i}}{a^{3/2}(t)} \sin\left[ m_i t + \theta_i \right]
\end{equation}
similarly to the single field case \cite{KLS97}.  The coefficients $\phi_{0,i}$ and phases $\theta_i$ are determined by the initial conditions at the end of N-flation.  The scale factor, averaged
over several oscillations, grows as $a(t) \sim t^{2/3}$ while the energy density in the N-flaton fields decreases as $\sum_i \rho_{\phi_i} = \sum_i\left[\frac{\dot{\phi}_i^2}{2} + \frac{m_i^2}{2}\phi_i^2\right] \sim a^{-3}$.
Hence, the coherent oscillations of the inflatons after N-flation behave very much as massive, non-relativistic particles. 

The homogeneous dynamics of the N-flatons, including the phase of post-inflationary oscillations, were studied in detail in \cite{multi_preheat}.  For reasonable choices of mass spectrum and initial conditions (at the beginning of inflation)
it was shown that, at the onset of the phase of oscillations, more than 90 percent of the energy is confined to the lightest 10 percent of the fields in a very narrow mass range.   
The remaining heavier fields will have already settled down at the minimum by this stage. Since the fields which survive to the end of inflation are nearly degenerate in mass, their post-inflationary oscillations are nearly in phase; see Fig.~\ref{fig:field1500}. 
It is this subset of fields that plays an important dynamical role during (p)reheating.  More on this later.

\subsection{String Theory Realization}

In \cite{Nflation} it was argued that the potential (\ref{V_i}) is radiatively stable.  However, since the flatness of the inflaton potential may be sensitive to Planck-suppressed corrections, it is useful to 
comment on how this scenario might arise in string theory.  (See also \cite{KSS} and \cite{grimm}.)  For the sake of being concrete, suppose we consider type IIB string theory and stabilize the moduli using the method of 
KKLT \cite{kklt}.  (Similar results are expected in the stable IIA vacua of \cite{IIA}.)  In the dimensional reduction, there arise $N = h_{1,1}$ axions $\tilde{\phi}_i$ associated with the 4-cycles of the Calabi-Yau manifold.\footnote{The meaning
of the tilde notation $\tilde{\phi}_i$ will become clear shortly.}  In a supersymmetric compactification, these axions combine with the volumes of the 4-cycles $\tau_i$ to form the complex parameter  $T_i = \tau_i - i \tilde{\phi}_i$.  
The KKLT super-potential is $W = W_0 + \sum_i A_i e^{-a_i T_i}$ where $W_0$, $A_i$ depend on the dilaton and complex structure.  
The exponential terms in the super-potential arise due to nonperturbative effects (one for each 4-cycle).
To leading order, the Kahler potential can be written as $K = -2 \ln(\mathcal{V})$ where the Calabi-Yau volume $\mathcal{V}$ depends on
the 4-cycle moduli $\tau_i$ in a non-trivial way.  For small values of the axions $\tilde{\phi}_i$, the action takes the schematic form
\begin{equation}
\label{non_can}
  \mathcal{L} = -\frac{1}{2}G_{ij}\partial_\mu \tilde{\phi}^i \partial^\nu \tilde{\phi}^j - \frac{(M^2)_{ij}}{2}\tilde{\phi}^i\tilde{\phi}^j + \cdots  \,\, .
\end{equation}
In general, neither the mass matrix $(M^2)_{ij}$ nor the metric on field space $G_{ij}$ is diagonal.  To put (\ref{non_can}) into canonical form we must perform a rotation in field space.  
In \cite{matrix_spectrum} this diagonalization was performed explicitly.  For our purposes, it suffices simply to note that one must generically perform a transformation of the form
\begin{equation}
\label{field_rot}
  \tilde{\phi}_i = \sum_{j} a_{j}^{(i)} \phi_j
\end{equation}
to put (\ref{non_can}) in the canonical form with N-flaton potentials given by (\ref{V_i}).  In \cite{matrix_spectrum} it was shown that the masses $m_i$ are distributed according to the Marcenko-Pastur
law, independently of the microscopic details of the compactification.  This distribution, which we assume throughout this work, depends on a single parameter $\beta$ which counts the number of axions divided by the dimension of the moduli space.
Where explicit numerical values are required, we assume the value $\beta = 1/2$, which was advocated in \cite{matrix_spectrum}.

An important observation is that the interactions between the N-flatons and visible matter will typically be diagonal in the basis $\tilde{\phi}_i$ which is associated with the geometrical 4-cycles of the 
Calabi-Yau manifold.  For example, we expect a coupling of the form $\tilde{\phi}_j F \wedge F$ to gauge fields living on a D7-brane wrapping the $j$-th cycle of the Calabi-Yau manifold.  As a result of the field rotation (\ref{field_rot}),
we generically expect that \emph{all} canonical N-flatons couple to each visible matter sector.

\section{Preheating After N-flation}
\label{sec:preheating}

\subsection{Couplings to Scalar Field Matter}
\label{subsec:scalar_couplings}

In this section we study the possibility of strong nonperturbative preheating effects after N-flation.  To proceed, we must first couple the inflatons to some matter fields.  For simplicity, we first model matter by a single light scalar field $\chi$.  
(Later on we will consider more realistic couplings to brane-bound gauge fields.)  Since we do not expect the interaction basis $\tilde{\phi}_i$ and the mass basis $\phi_i$ to be aligned, we should consider couplings between $\chi$ and \emph{all} 
the N-flatons.  We therefore propose a prototype Lagrangian of the form
\begin{eqnarray}
  \mathcal{L} &=& \sum_{i=1}^N \left[ -\frac{1}{2}\partial_\mu \phi_i \partial^\mu \phi_i - \frac{m_i^2}{2}\phi_i^2  \right] \nonumber \\
              && - \frac{1}{2}\partial_\mu \chi \partial^\mu \chi - \sum_{i=1}^N \left[ \frac{g_i^2}{2} \phi_i^2 + \frac{\sigma_i}{2}\phi_i \right] 
                      \chi^2 - \frac{\lambda}{4}\chi^4 \label{L2} \, .
\end{eqnarray}
The term $\lambda \chi^4$ is required to keep the potential bounded below, but otherwise will play no role in our discussion.  Treating the N-flatons as classical background fields (\ref{Nosc}), the equation
of motion for the linear quantum fluctuations of the $\chi$ field takes the form
\begin{equation}
\label{gen_eom}
  \ddot{\chi} + 3 H \dot{\chi} + \frac{\nabla^2}{a^2} \chi + \sum_i\left[ g_{i}^2 \phi_i^2(t) + \sigma_i \phi_i(t)  \right]\chi = 0 \, .
\end{equation}
We introduce the variable $X(t,{\bf x}) =a^{3/2}(t) \chi(t,{\bf x})$ and write the equation of motion (\ref{gen_eom}) in the form 
\begin{equation}
\label{X_4}
  \ddot{X}_k(t) + \left[\frac{k^2}{a^2} + \sum_i\frac{g_i^2 \phi_{0,i}^2}{a^3(t)}\sin^2\left(m_i t + \theta_i\right) + \sum_i \frac{\sigma_i\phi_{0,i}}{a^{3/2}(t)}\sin\left(m_i t + \theta_i \right) + \Delta \right] X_k(t) = 0 \, ,
\end{equation}
where $\Delta = -\frac{3}{4}\left(2\frac{\ddot{a}}{a}+H^2\right)=\frac{3}{4}\frac{P}{M_p^2}$ with $P$ the pressure of the dominant inflaton background.  Since the N-flaton oscillations (\ref{Nosc}) behave as non-relativistic matter,
$P\cong 0$ and we can ignore the quantity $\Delta$ in (\ref{X_4}).  The effective mass for the $X_k$ modes is
\begin{equation}
\label{Meff}
  M_{\mathrm{eff}}^2(t) = \sum_i\frac{g_i^2 \phi_{0,i}^2}{a^3(t)}\sin^2\left(m_i t + \theta_i\right) + \sum_i \frac{\sigma_i\phi_{0,i}}{a^{3/2}(t)}\sin\left(m_i t + \theta_i \right) \, .
\end{equation}
If the N-flaton masses were precisely degenerate, $m_i \equiv m$, then the oscillatory contributions to (\ref{Meff}) would add coherently, mimicking the well-studied single-field case \cite{KLS97,tac_res}.  More realistically, however,
the inflaton masses $m_i$ should be sampled from some distribution such as the Marcenko-Pastur law \cite{matrix_spectrum}.  In this case the time-dependent behaviour of the effective mass (\ref{Meff}) may be quite different from the single-field
case and, therefore, preheating may proceed differently.  To understand the behaviour of the solutions of (\ref{X_4}) we should first study $M_{\mathrm{eff}}^2(t)$ for parameter choices that are reasonable for N-flation.  In passing,
notice that the summation over N-flaton field indices in (\ref{X_4}) and (\ref{Meff}) need only include the lightest N-flatons which remain dynamical until the onset of post-inflationary oscillations.

In constructing solutions of (\ref{X_4}) there are two limiting cases which may be of interest, depending on which interaction in (\ref{L2}) dominates.  It may happen that the 4-leg interactions $\sum_i g_i^2 \phi_i^2 \chi^2$ 
dominate over the 3-leg interactions $\sum_i\sigma_i \phi_i\chi^2$, or vise versa.  Let us first study the behaviour of $M_{\mathrm{eff}}^2(t)$ in each limiting case, to demonstrate the absence of strong parametric resonance preheating
via the 4-leg interactions and the possibility of strong tachyonic resonance preheating via the 3-leg interactions.  Later on, we will confirm these results by developing the mathematical theory of the quasi-periodic Mathieu equation.  In that case, we will
see that \emph{both} limiting cases can be treated simultaneously by means of a simple device.

\subsection{Four-Leg Interactions: Absence of Strong Adiabaticity Violations}

Let us first study the time dependence of the effective mass (\ref{Meff}) in the limit where the 3-leg interactions are vanishing, that is $\sigma_i=0$.  Now (\ref{Meff}) takes the form
\begin{equation}
\label{M4legs}
  M_{\mathrm{eff}}^2(t) \cong \sum_i g_i^2 \phi_i^2(t) = \sum_i\frac{g_i^2 \phi_{0,i}^2}{a^3(t)}\sin^2\left(m_i t + \theta_i\right) \, .
\end{equation}
In Fig.~\ref{fig:quadcoupling} we illustrate numerically the typical time-dependence of the effective mass (\ref{M4legs}) for $N=600$ inflatons with the Marcenko-Pastur
mass distribution \cite{matrix_spectrum} and randomly distributed initial conditions $\phi_{i}(t_0)$ at the beginning of N-flation (the qualitative behaviour is independent 
of this specific choice).

\DOUBLEFIGURE[ht]{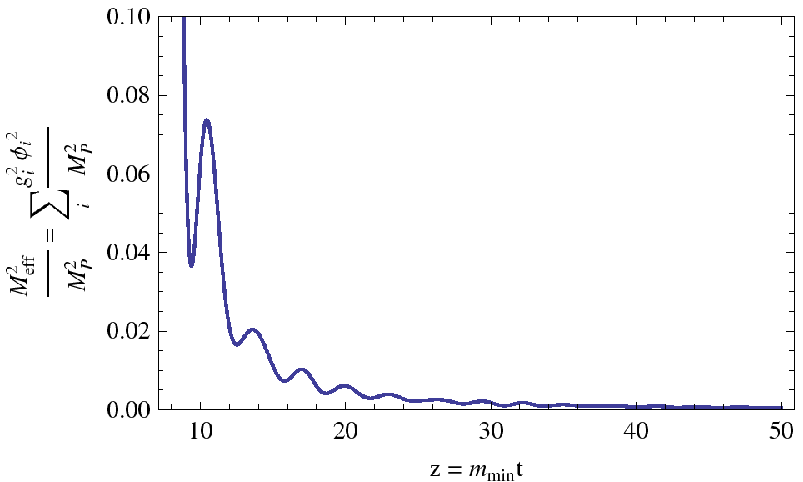,width=2.5in,height=1.5in}{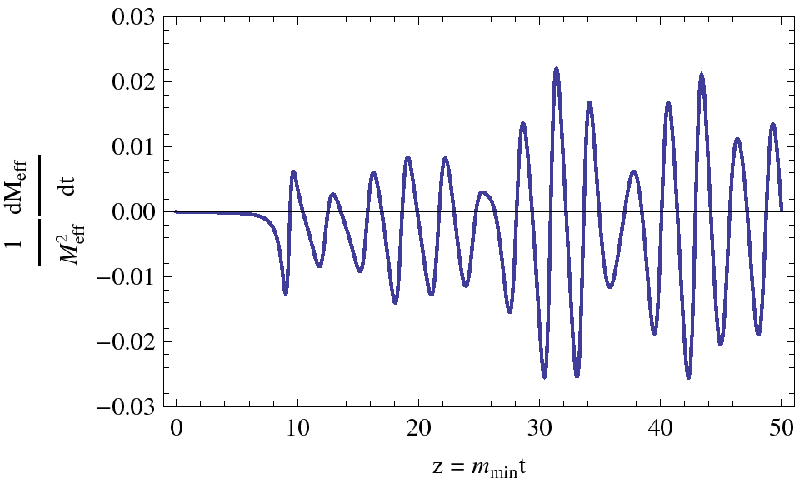,width=2.5in,height=1.5in}{Evolution of the effective mass $M_{\mathrm{eff}}^2(t)=\sum g_i^2\phi_i^2(t)$ of the $\chi$
field for $N=600$, assuming randomly distributed initial field values at the start of inflation.
Only the oscillatory part of the motion relevant for reheating has been illustrated.
For simplicity, we have set all the coupling constants $g_i=1$.
\label{fig:quadcoupling}}{The time evolution of the quantity $\dot{M}_{\mathrm{eff}} / M_{\mathrm{eff}}^2$
for the effective mass $M^2_{\mathrm{eff}}(t) = \sum_ig_i^2 \phi_i^2(t)$ and the same parameters as Fig.~1.
This quantity is a measure of adiabaticity of the mode functions.  For parameter choices relevant for N-flation strong violations of adiabaticity are absent.\label{fig:quadadiabatic}}

The spread of the masses $m_i$ in a realistic N-flation model prevents the effective mass (\ref{M4legs}) from crossing zero and impedes the violations of adiabaticity that are associated with broad band
parametric resonance.  To illustrate this we plot the quantity $\dot{M}_{\mathrm{eff}} / M_{\mathrm{eff}}^2$, which measures the adiabaticity of the mode functions, in Fig.~\ref{fig:quadadiabatic}.  We have verified
the smallness of the adiabaticity parameter in the large-$N$ limit for several choices of initial conditions (see also \cite{multi_preheat}).  Therefore, we do not expect any strong nonperturbative preheating
effects for the 4-leg interactions $\sum_i g_i^2 \phi_i^2 \chi^2$, assuming $N\gg 1$ and the Marcenko-Pastur mass spectrum.  This expectation was verified numerically in \cite{multi_preheat}.  Below, we confirm 
these results using the mathematical theory of the quasi-periodic Mathieu equation.

\subsection{Three-Leg Interactions: Possibility of Tachyonic Resonance}

Let us now consider the opposite regime, wherein the 3-leg interactions in (\ref{L2}) dominate.  We therefore seek to solve the mode equation (\ref{X_4}) with $g_i^2= 0$. In this limiting case the effective mass 
(\ref{Meff}) takes the form
\begin{equation}
\label{M3legs}
  M_{\mathrm{eff}}^2(t) = \sum_i \sigma_i \phi_i(t) \cong \sum_i\frac{\sigma_i \phi_{0,i}}{a^{3/2}}\sin\left(m_i t + \theta_i\right) \, .
\end{equation}
The resulting mode equation is
\begin{equation}
\label{X_3}
  \ddot{X}_k(t) + \left[\frac{k^2}{a^2} + \sum_i\frac{\sigma_i \phi_{0,i}}{a^{3/2}}\sin\left(m_i t + \theta_i\right) + \Delta \right] X_k(t) = 0 \, ,
\end{equation}
where $\Delta = \frac{3}{4}\frac{P}{M_p^2} \cong 0$ can be neglected. The behaviour of the effective mass (\ref{M3legs}) is \emph{very} different from the 4-legs case, equation (\ref{M4legs}).  In the presence of 
dominant 3-leg interactions the effective mass-squared can go tachyonic, $M_{\mathrm{eff}}^2(t) < 0$, at certain stages during the N-flaton oscillations.  In the single field case, this leads to the well-studied 
tachyonic resonance preheating \cite{tac_res}.  During a tachyonic phase the fluctuations $X_k(t)$ will display exponential growth, which is associated with the explosive production of $\chi$ particles by spinodal 
instability. 

\begin{figure}[h!]
  \centering
    \includegraphics[width=0.5\textwidth]{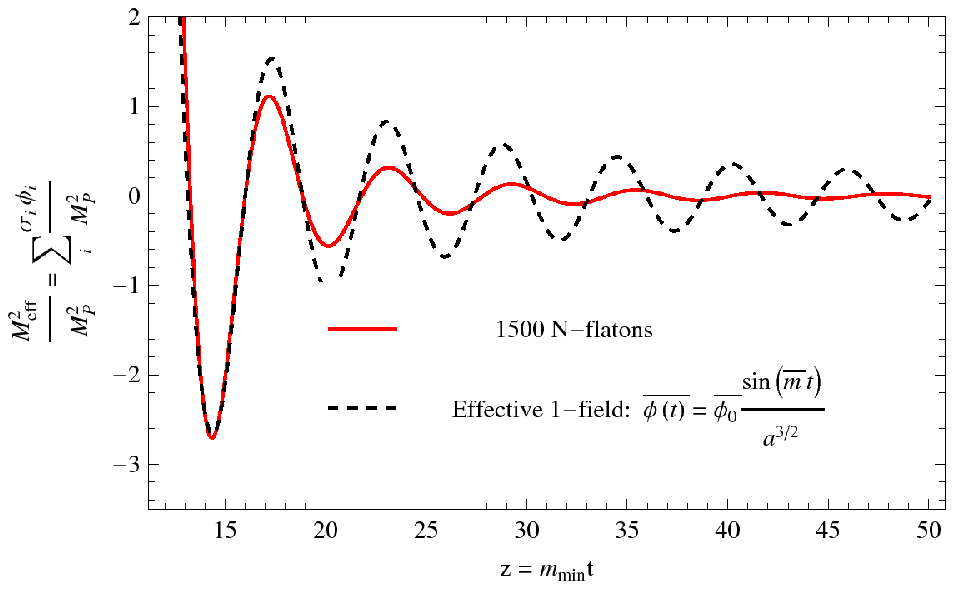}
  \caption{The post-inflationary time-dependence of the effective mass $M^2_{\mathrm{eff}}(t) = \sum_i\sigma_i\phi_i(t)$ for the 3-leg interaction.For simplicity, we assume $\sigma_i=\sigma$.  This figure shows that the post-inflationary oscillations in 
  $M^2_{\mathrm{eff}}(t)$ are qualitatively identical to what one would obtain for single-field inflation (corresponding to the $\sin(\bar{m}t)$ solution), 
  aside from a slowly time-dependent frequency and more rapid damping of the amplitude.  In plotting the analytic 
  solution, we have numerically matched the amplitudes at the first mimima and computed $\bar{m}$ from the first half oscillation of the multi-field solution.\label{fig:freq1500}}
\end{figure}

The strength of the nonperturbative tachyonic resonance effects associated with (\ref{X_3}) depend sensitively on the couplings $\sigma_i$ and the initial conditions at the onset
of reheating.  As discussed previously, the heaviest N-flatons drop out of the dynamics rapidly during N-flation.  During the post-inflationary oscillations, the majority of the energy is carried by
the lightest 10 percent of the N-flatons with a very narrow range of masses.  Since these N-flatons have nearly degenerate mass, they start to oscillate at roughly the same time, leading to a picture where
the oscillations (\ref{Nosc}) are more-or-less in phase, as illustrated in Fig.~\ref{fig:field1500}.  This approximate coherence results in a time-dependence for the effective mass (\ref{M3legs}) which is nearly identical to the results one would obtain
for a single field inflation model.\footnote{As we saw above, in the case of the 4-leg interactions (\ref{M4legs}) any slight de-phasing of the N-flaton oscillations makes violations of adiabaticity exceedingly unlikely
in the limit $N\gg 1$.  This occurs because (\ref{M4legs}) is a sum of $N$ positive definite terms, when $N$ is large zero-crossings become very rare.  However, in the case of the 3-leg interactions the effective mass (\ref{M3legs})
is very different; it is a sum of $N$ terms with indefinite sign.  Hence, small de-phasing of the N-flaton oscillations do not conspire to forbid tachyonic regions.}  The resulting dynamics of the effective mass (\ref{M3legs}) 
are illustrated in Figure~\ref{fig:freq1500}.

From Fig.~\ref{fig:freq1500} we see that the effective mass (\ref{M3legs}) looks like that of a single field with a slowly time-varying frequency
and an amplitude damping faster than would be expected from the expansion alone.  Both of these effects are due to the de-phasing of the various 
N-flaton fields (this may be verified analytically also).

In summary: we have seen that a realistic choice of N-flationary parameters leads to a behaviour of the effective mass (\ref{M3legs}) in the case of 3-legs interactions which is nearly identical to the usual single-field result.  Hence, preheating
of the $X_k$ fluctuations will proceed via the same tachyonic resonance mechanism which was studied in detail in \cite{tac_res,ref:tacres2}.  We do not repeat this analysis here.  On the other hand, we can imagine a scenario wherein a coupling of
the form (\ref{M3legs}) is relevant for preheating, however, the various inflaton fields have amplitudes which are more-or-less randomly distributed at the onset of \emph{reheating}.\footnote{This is, of course, very different from assuming randomly
distributed inflaton field values at the onset of \emph{inflation}.}  For example, such a situation might arise in multi-field models where inflation proceeds via some extremely complicated trajectory through the string landscape 
\cite{multi_land,staggered,rand1,rand2,stream}.  Therefore, let us briefly consider a toy scenario (not necessarily relevant for N-flation) where $N=5$ inflatons have random initial displacements satisfying $0.2 M_p \leq |\phi_{0,i}| \leq M_p$ at the 
onset of reheating.  We also assume a random mass spectrum.  Our choice of parameters for this example is given explicitly in Table \ref{table:5fieldic}.  By choosing a somewhat large hierarchy between the heaviest and lightest field we can obtain results 
which are qualitatively different from the single-field case.

\begin{table}[h]
\caption{Masses and Initial Displacements of Inflatons}
\centering
  \begin{tabular}{c c c}
  \hline\hline
  $i$ & $\left(\frac{m_i}{m_{\mathrm{min}}}\right)^2$ & $\frac{\phi_{0,i}}{M_P}$ \\ [1ex]  
  \hline
1 & 2.43 & 0.533 \\
2 & 5.63 & 0.412 \\
3 & 9.88 & 0.739 \\
4 & 15.85 & 0.544 \\
5 & 26.67 & 0.701 \\ [0.5ex]
\hline
\end{tabular}
\label{table:5fieldic}
\end{table}

In Fig.~\ref{fig:inflatonbg} we display the post-inflationary oscillations of the heaviest and lightest inflaton fields for the example parameters in Table \ref{table:5fieldic}.  
In Fig.~\ref{fig:trilincoup} we plot the time-dependence of the effective mass-squared (\ref{M3legs}) for this same example, showing the presence of tachyonic regions where $M_{\mathrm{eff}}^2(t)$. 
These regions leads to violent particle production via tachyonic resonance.  From Fig.~\ref{fig:trilincoup} it is also clear the behaviour of $M^2_{\mathrm{eff}}(t)$ is qualitatively different 
from the single field case (compare with Fig.~\ref{fig:freq1500}).

\DOUBLEFIGURE[ht]{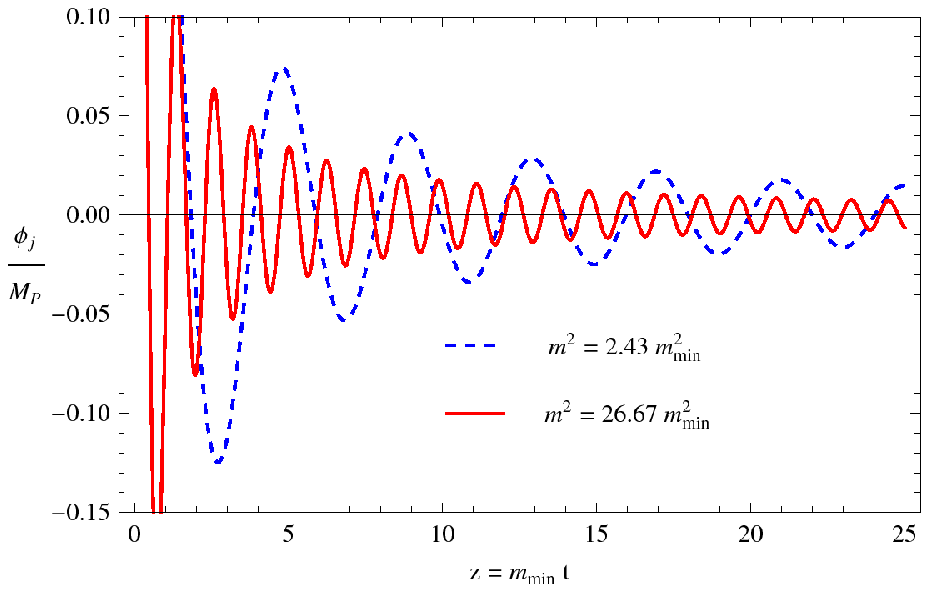,width=2.25in}{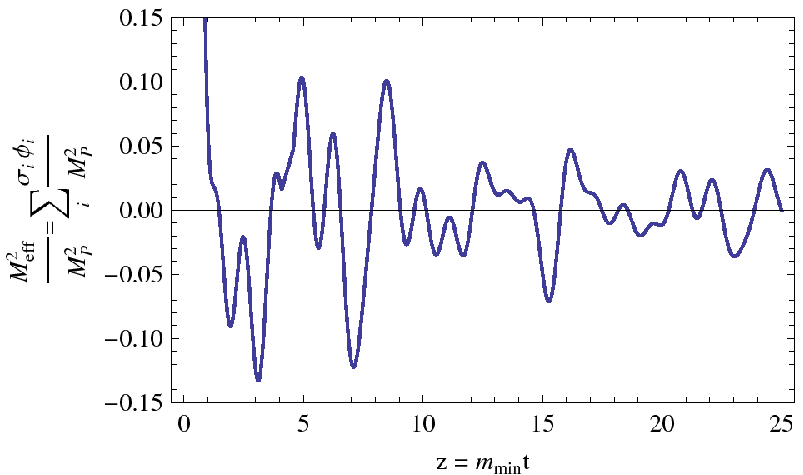,width=2.25in}{The evolution of the most massive (solid) and least massive (dashed) background inflatons for five fields for the parameters given in Table 1.  
As we see, each inflaton behaves as a damped harmonic oscillator as expected.  As usual, the fields values are in units of $M_p$ and the time is $z = m_{\mathrm{min}}t$.\label{fig:inflatonbg}}{The evolution 
of the effective mass $M_{eff}^2 = \sum_i\sigma_i\phi_i(t)$ of the $\chi$ field for the example in Table 1.  We assume that all couplings are equal: $\sigma_i = \sigma$ for all $i$.
This figure illustrate the presence of multiple tachyonic regions during the inflaton oscillations.  The dynamics of the effective mass is very far from periodic, which implies 
that this example is qualitatively different from the single-field case.\label{fig:trilincoup}}

The tachyonic regions illustrated in Fig.~\ref{fig:trilincoup} and the non-periodic oscillations of the effective frequency lead to unstable growth of the fluctuations $X_k(t)$ by tachyonic resonance.  
The exponential instability of the $\chi_k$ inhomogeneities is associated with explosive particle production.  In Fig.~\ref{fig:k40chi} and Fig.~\ref{fig:stablek_field} we have illustrated the unstable behaviour of
the solutions $X_k(t)$ of (\ref{X_3}) taking, again, the parameter values in Table \ref{table:5fieldic}.  These figures illustrate the rapid growth of inhomogeneities, showing that this new type of multi-field tachyonic
resonance is very efficient.  Within a very short time, back-reaction effects become important and one must turn to lattice simulations to fully capture the dynamics.

\DOUBLEFIGURE[ht]{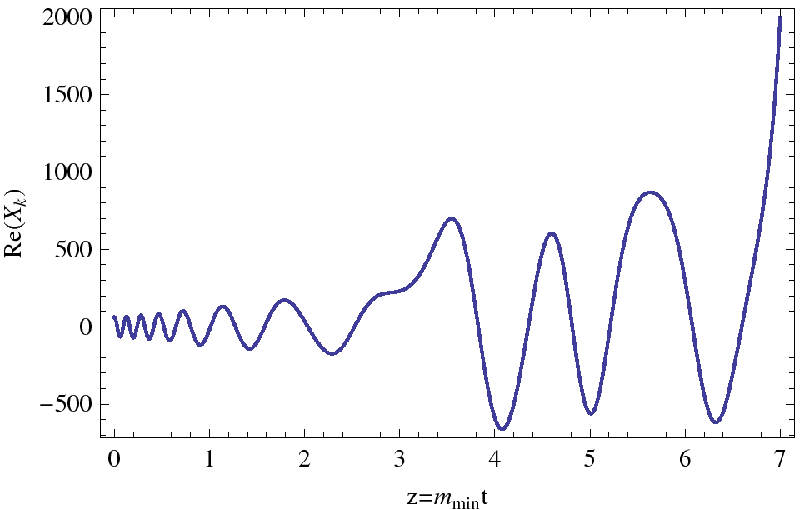,width=2.25in}{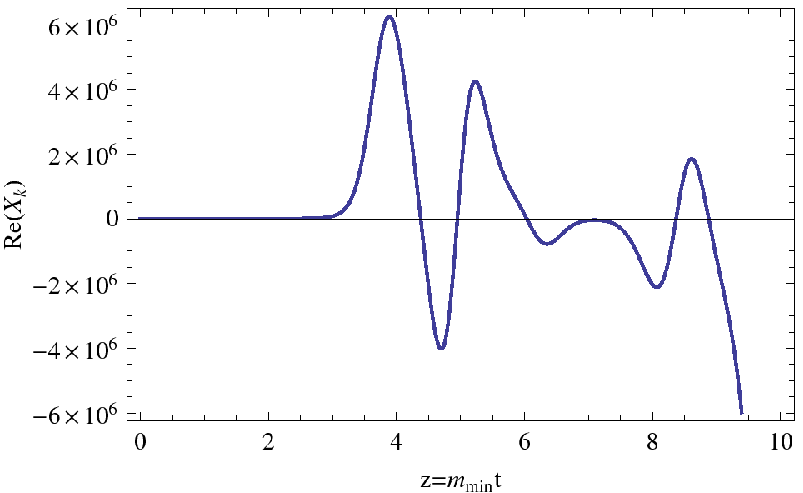,width=2.25in}{The unstable growth of the quantum field $X_{\bf k}(t)$ due to multi-field tachyonic resonance effects.  For illustration we take 
$k=40m_{\mathrm{min}}$.\label{fig:k40chi}}{The 
unstable growth of the quantum field $X_{\bf k}(t)$ due to multi-field tachyonic resonance effects.  For illustration we take $k=9.608m_{\mathrm{min}}$.\label{fig:stablek_field}}

\subsection{The Quasi-Periodic Mathieu Equation}
\label{subsec:QP}

We have argued that strong nonperturbative effects are absent for the 4-leg interactions (\ref{M4legs}), whereas tachyonic resonance may proceed in the case of the 3-leg interactions (\ref{M3legs}).
We will now re-consider these results using the mathematical theory of the quasi-periodic Mathieu equation~\cite{rrand1,rrand2,restrap,broer1}.  Using this perspective, we can unify the discussion of both the 4-legs and 3-legs interactions in (\ref{L2}).  
To see this, note that, in the limit where the 4-leg interactions dominate, equation (\ref{X_4}) yields
\begin{equation}
\label{4-thingy}
  \ddot{X}_k(t) + \left[\frac{k^2}{a^2} + \sum_i\frac{g_i^2 \phi_{0,i}^2}{a^3(t)}\sin^2\left(m_i t + \theta_i\right) + \Delta \right] X_k(t) \cong 0 \, .
\end{equation}
On the other hand, in the limit where the 3-legs interactions dominate, we would have
\begin{equation}
\label{3-thingy}
  \ddot{X}_k(t) + \left[\frac{k^2}{a^2} + \sum_i\frac{\sigma_i \phi_{0,i}}{a^{3/2}(t)}\sin\left(m_i t + \theta_i\right) + \Delta \right] X_k(t) \cong 0 \, .
\end{equation}
In \emph{both} cases, if we ignore the expansion of the universe, an appropriate redefinition of our time parameter puts the equation in the form
\begin{equation}
  \label{hill_3}
  \frac{d^2}{dz^2} X_k(z) + \left[A_k - 2 \sum_{i=1}^{N_{\mathrm{osc}}} q_i \cos\left( 2 \frac{m_i}{m} z + \theta_i \right)  \right] X_k(z) = 0 \, ,
\end{equation}
where $m$ is some convenient mass scale (such as the mass of the heaviest or lightest inflaton, or the average mass of the inflatons) and $N_{\mathrm{osc}}$ is the number of N-flatons which play a significant
role in reheating (that is, $N_{\mathrm{osc}}$ counts only those lightest N-flatons which oscillate at the end of inflation).  Therefore, we can present of unified treatment of both limiting cases by developing the theory of 
equation (\ref{hill_3}).

Let us now discuss the solutions of the quasi-periodic Mathieu equation (\ref{hill_3}).  In the simplest case of a single inflaton field with $m_i = m$ and $q_i = q$.  Then, neglecting the expansion of the universe, equation (\ref{hill_3}) reduces to the
Mathieu equation \cite{ref:Mathieu} which is well-known from studies of preheating after inflation \cite{KLS97}.  This equation is known to posess instability bands, for certain values of $A_k$ and $q$ the solutions display exponential growth
\begin{equation}
\label{unstable_grow}
  X_k(z) = e^{\mu_k z} P_k(z) \, ,
\end{equation}
where $P_k(z)$ is some periodic function and the real number $\mu_k$ is the Floquet exponent which determines the rate of the instability.  The stability chart of the Mathieu equation, depicted in the 
left panel of Figure~\ref{fig:stabchart}, can be used to explain the resonance structure of preheating for single field inflation~\cite{KLS94,KLS97}.  The shaded regions of this chart display parameter values
for which the solutions of (\ref{hill_3}) have Floquet exponent $\mu_k \not= 0$ and hence display unstable growth.

For the case at hand, multiple inflaton fields, there are two cases to distinguish depending on the values of the oscillation frequencies $m_i / m$: either these are all rationally related or else at least two of them are incommensurate.
In the first case, the effective frequency for $X_k$ is periodic and Floquet's theorem ensures that for certain values of the parameters $A_k$ and $q_i$ the solutions will again grow as (\ref{unstable_grow}).
Instead, if two of the masses are irrationally related, then the effective frequency is no longer periodic and Floquet's theorem no longer applies.  When this occurs, (\ref{hill_3}) is known as the quasi-periodic (QP) Mathieu equation.
We will study the theory of this equation below.  With some abuse of language, we will refer to~\eqref{hill_3} as the QP Mathieu equation and the exponential growth rates as the Floquet exponents regardless of the actual mass spectrum.
One may worry that the growth is no longer exponential in the QP case.
However, since an infinitesimal change in the masses can change their ratios from commensurate to incommensurate and vice versa, 
we expect exponential growth in this case also.

In the QP case (incommensurate frequencies), we can also make a further distinction: all of the masses and their sums and differences are incommensurate,
or at least one pair of masses (or some pair of sums or differences) are rationally related.  The second of these can lead to the phenomenon of resonance trapping 
~\cite{restrap} and may play a role when both 3-legs and 4-legs interactions are relevant during preheating.
A detailed study of these effects is beyond the scope of this paper and we will instead concentrate on cases where only 
a single type of interaction is relevant during preheating.

So far, we have ignored the expansion of the universe.  Adding the expansion makes the resonance parameters $A_k$ and $q_i$ time dependent (see below), and the results of Floquet's theorem will no longer apply, even for rationally related masses.
However, notice that $H \ll m_i$ during reheating, thus the time scale for variation of $a(t)$ is much longer than the period of oscillations in the effective frequency.  We can therefore employ an adiabatic approximation wherein
we solve (\ref{hill_3}) using the standard techniques but allow $A_k$, $q_i$ to vary slowly in the solution as a result of the expansion of the universe.  This leads to a picture where the
dynamics of $a(t)$ causes the resonance parameters $q_i$ to ``flow'' across the stability/instability chart.  
For a similar discussion in the single field case see~\cite{ref:steinhardt}.
More details on this shortly.

\subsection{Stability Chart and Floquet Exponents}
\label{subsec:features}

Having given the general picture of the relation between the QP Mathieu equation and reheating, 
we now turn to develop a mathematical understanding of the stability/instability chart associated with (\ref{hill_3}) 
in the interesting multi-field case.  Later on, we will apply these results to the specific reheating models under consideration.

In the case $N_{\mathrm{osc}} > 1$ the equation (\ref{hill_3}) is \emph{immensely} more complicated than the usual Mathieu equation.
As a warm-up, we first consider the illustrative case $N_{\mathrm{osc}} = 2$ with $q_1=q_2 \equiv q$ and neglect the phases $\theta_1 = \theta_2 = 0$.
(This very restrictive case may actually have some relevance for N-flation since only a relatively small number of N-flatons with nearly
degenerate masses are expected to participate in preheating.)  For this simple example we have solved~\eqref{hill_3} numerically for a wide 
range of parameters $A_k$ and $q$, taking $m_1/m = 1$ and $m_2/m = 0.3475$.  We have numerically constructed the stability/instability chart,
and the result is displayed in the right panel of Figure~\ref{fig:stabchart}.  

\EPSFIGURE[ht]{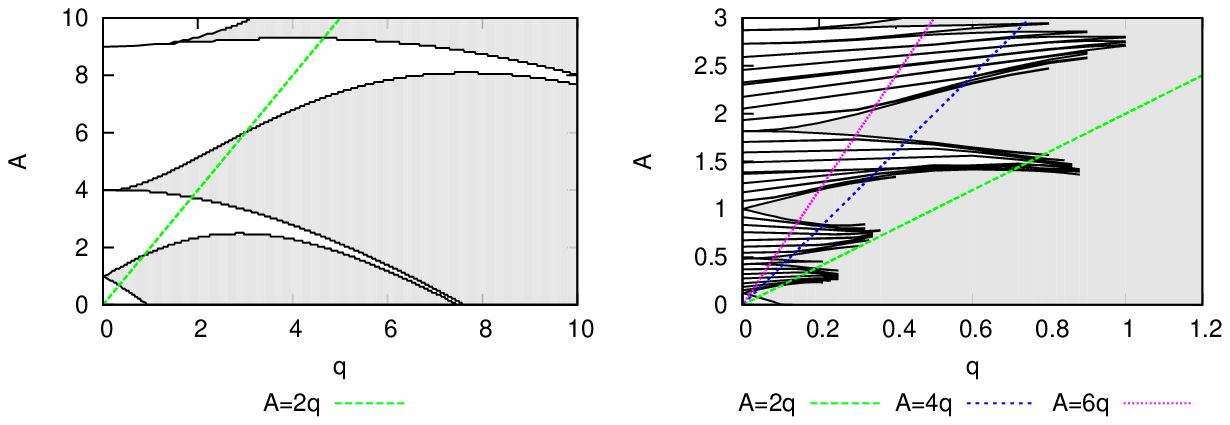}{In the left panel we show the stability/instability chart for the usual Mathieu equation, that is (\ref{hill_3}) with $N_{\mathrm{osc}} = 1$, $m_1=m$
and $q_1=q$.  In the right panel, we display the analogous stability/instability chart for the quasi-periodic Mathieu equation, that is (\ref{hill_3}) with $N_{\mathrm{osc}} = 2$,
$m_2 = 0.3475 m_1$ and $q_1=q_2\equiv q$.  In \emph{both} panels, the white regions of parameter space correspond to stable oscillatory solutions, whereas the shaded regions correspond to parameter values
for which the solutions of (\ref{hill_3}) display unstable, exponential growth.  We see that the inclusion of an additional oscillatory contribution to the effective frequency of equation
(\ref{hill_3}) leads to the formation of instability ``tongues'' which permeate the regions that were formerly associated with stability.  There are, in fact, an infinite number of such 
instability tongues suggesting that the stability region might be dissolved.  However, we only display a finite number of instability tongues, neglecting higher order effects which lead to
\emph{very} weak growth that cannot be accurately captured by finite-duration numerical simulations.
The relevance of the lines $A=2q$, $A=4q$ and $A=6q$ is discussed below.
\label{fig:stabchart}}

As can be seen from Fig.~\ref{fig:stabchart}, the non-commensurate ``extra'' oscillatory contribution to (\ref{hill_3}) leads to the formation of ``new'' instability regions at small $q_i$, called Arnold tongues.
These emerge from the $A$-axis with locations given by
\begin{equation}
  A = \left[\sum_i n_i \frac{m_i}{m} \right]^2 \, ,
\end{equation}
where each $n_i$ is an integer and the sum runs over the oscillating inflatons~\cite{rrand1,rrand2}.
We will refer to $\sum_i |n_i|$ as the order of the resonance.
For $q_i \ll 1$, the tongue labeled by the vector ${\bf n} = (n_1,...,n_{N_{\mathrm{osc}}})$
has width $\sim \prod_i|q_i|^{|n_i|}$.
If the masses are all rationally related, then eventually the higher order resonances
will be located at the same $A$ value as a lower order resonance and the number of
Arnold tongues will be finite.
However, if we have incommensurate masses, then the tips of these instability tongues
will densely fill the $A$ axis.  Thus, one might naively expect that in this case all 
of the modes will be unstable and the resonance will be very efficient.  However, even if the stability
bands completely dissolve, this does not necessarily mean that preheating will become significantly more efficient.  
It might be that the Floquet exponents $\mu$ in the dissolved regions are very small.

Let us now discuss the Floquet exponents $\mu$ associated with the quasi-periodic Mathieu equation.  In order for the exponential growth (\ref{unstable_grow})
to be important, the Floquet exponent $\mu$ must be large enough that the growth $e^{\mu z}$ is appreciable over the time of reheating.  Moreover, the expansion
of the universe will cause modes to ``flow'' through the stability/instability chart, weakening parametric resonance effects as $a(t)$ grows.  If $\mu$ is not sufficiently
large, then there will be no important growth of the mode before it is red-shifted out of the instability band.  Let us consider the behaviour of the Floquet exponent in our
two-oscillator example with $N_{\mathrm{osc}} = 2$ with $q_1=q_2 = 0.05$ and $m_2=0.3475 m_1$.  Fig.~\ref{fig:floquetsmallq} shows $\mu$ as a function of $A$.  The two 
first order resonances ($\sum_i |n_i| = 1$) are much stronger than the second order ($\sum_i |n_i| = 2$) resonances, while the third order resonances are too weak to be visible.
Hence, in this case the fact that the Arnold tongues densely fill the $A$ axis in Fig.~\ref{fig:stabchart} turns out to be irrelevant: the ``instability'' associated with the higher order
tongues is \emph{far} too weak to be important for reheating.

It is interesting to compare the Floquet exponents for our example quasi-periodic oscillator to the usual Mathieu equation, in the regime $q \ll 1$.
From Figure~\ref{fig:floquetsmallq}, we see that for the resonances which are common to both the QP Mathieu and the periodic Mathieu equations (\emph{i.e.} those with only
a single $n_i$ non-zero), the Floquet exponents are nearly identical for both equations.  The effect of the multiple frequencies is simply to create new resonance bands
which do not overlap with those present in the periodic Mathieu equation.  Since these resonances are all of order 2 or greater, they are much weaker than
the first order resonances and are thus sub-dominant.

\EPSFIGURE[ht]{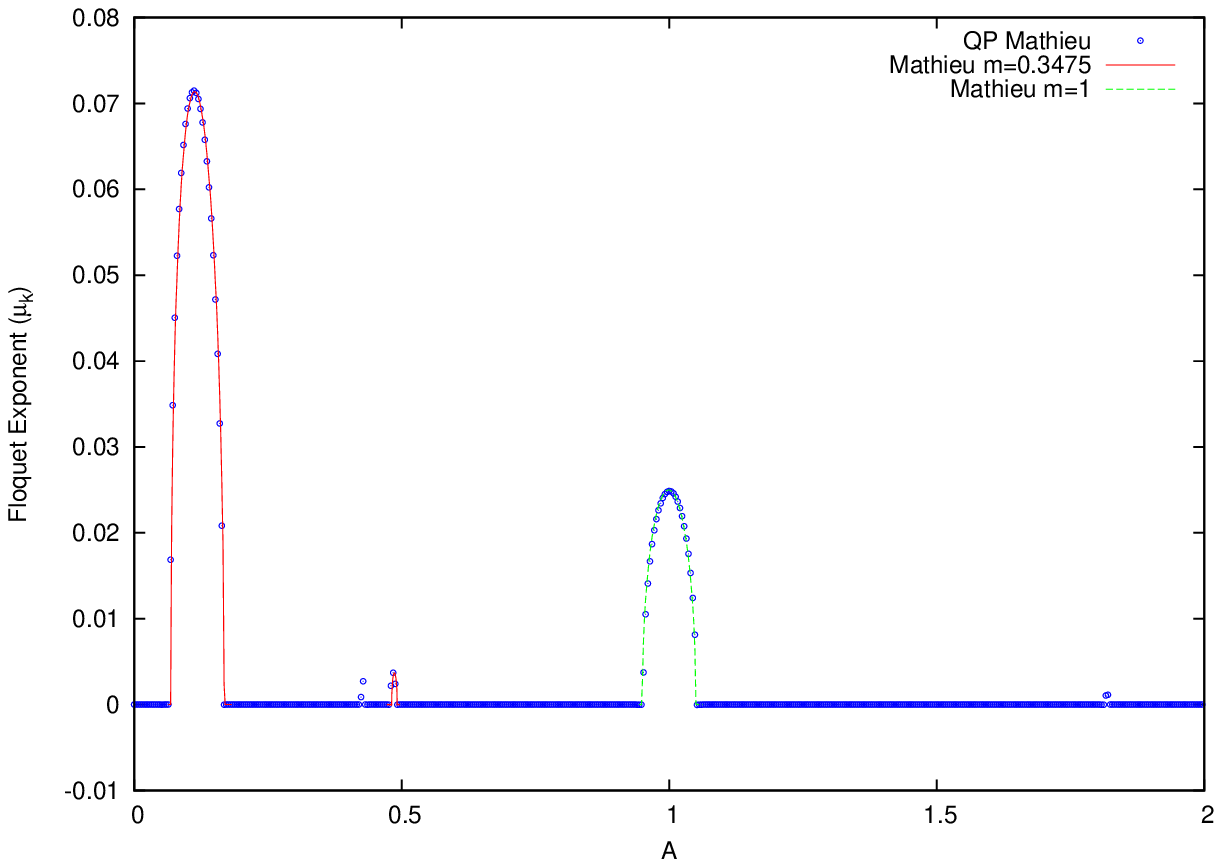,width=4in}{Floquet exponents for the case of $q_1=q_2=0.05$ and a range of
$A$ values.  The locations of the instability bands are clear, with
the two widest bands corresponding to the two first order resonances.
The remaining three resonances are all second order and correspond,
from left to right, to the $(n_1,n_2)=(1,-1)$, $(0,2)$ and $(1,1)$ resonances,
where the first index is for the $m_1=m$ oscillator and the
second for the $m_2=0.3475m$ oscillator.
Also shown are the Floquet exponents for the case of a single oscillator
(corresponding to the periodic Mathieu equation) with a mass $m_1 = m$ (dashed green curve) 
and $m_2 = 0.3475m$ (solid red curve).\label{fig:floquetsmallq}}


Thus far, we have focused on the Floquet exponents in the regime $q_i \ll 1$, which is associated with weak resonance.  In the single field case, $q\gg 1$
would be the condition for a strong parametric resonance.  How should this condition be generalized to the multi-field case?  If we denote the Floquet
exponents for the periodic Mathieu equation with $m_1 / m \equiv \omega$ as $\mu_{\omega}(A,q_1)$ then a simple re-definition $z \rightarrow \omega z$
shows that $\mu_{\omega}(A,q) = \mu_{\omega=1}(A/\omega^2,q/\omega^2) \omega$.  This demonstrates that the effective resonance parameter (when determining
Floquet exponents) for an oscillator with mass $m_i$ is given by $q^{\mathrm{eff}}_i \equiv q_i\left(m/m_i\right)^{2}$.
Hence, we expect (roughly) that strong resonance will occur in instability tongues involving the $i$-th oscillator (those with $n_i \neq 0$) 
whenever $q_i^{\mathrm{eff}} \gg 1$.
However, as we explain below, this intuition may need to be modified when $q$ becomes sufficiently large.

Consider now the approach to the strong-resonance regime of our toy two-oscillator example.  
From the stability diagram, we see that increasing $q$ causes the various resonance tongues to open up and begin to overlap.
This is a qualitatively new feature relative to the periodic case (where the tongues do not overlap).
We have illustrated the behaviour of the Floquet exponents in this overlap region for our two-oscillator example in Fig.~\ref{fig:floquetbigq}.
(We compare this two-oscillator case to the usual Mathieu equation with $m=1$ or $m=0.3475$.)
Prior to the overlap, the strength of the resonances can still be estimated by
considering the parameter $q_i^{\mathrm{eff}} = q_i(m/m_i)^2$ relevant for the single oscillator case.
However, once multiple oscillators contribute to the instability for a fixed value of $A$ and $q$,
the Floquet exponents become complicated functions of $A$ (for a fixed $q$).
When this occurs, it is much more difficult to determine the effective resonance parameter controlling the strength of the instability.

\EPSFIGURE[ht]{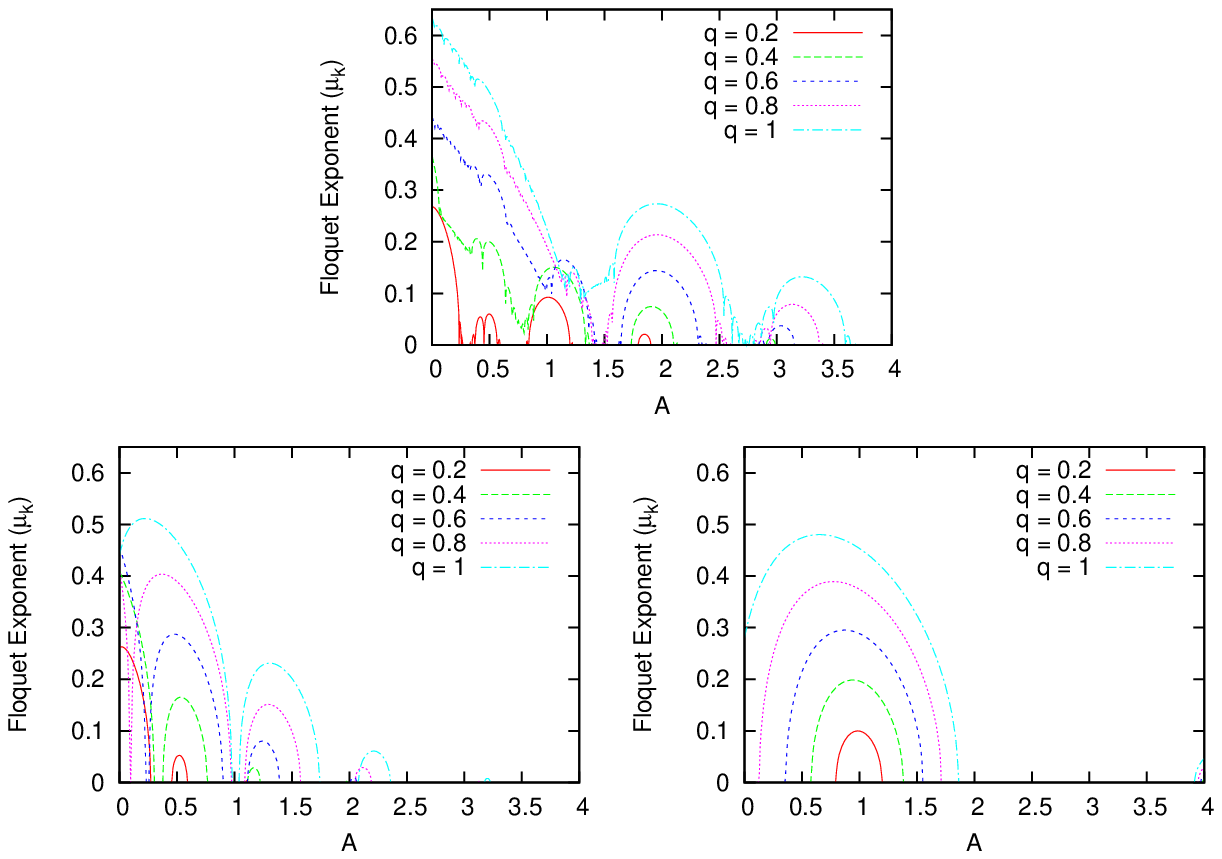}{Floquet exponents for several different equations as a
function of $A$ for a range of $q$ parameters.
In the upper panel are the exponents for the QP Mathieu equation
with $q_1 = q_2 = q$ and the same masses as in Figure~{\ref{fig:stabchart}}.
The bottom left panel is for the periodic Mathieu equation with $m = 0.3475$.
The bottom right panel is for the periodic Mathieu equation with $m = 1$.\label{fig:floquetbigq}}

Let us estimate the $q_i$ values at which the overlap displayed in Fig.~\ref{fig:floquetbigq} occurs.
Clearly, if we have $q_i^{\mathrm{eff}} \gsim 1$ for one or more of the oscillators, then there will be many
``open'' resonance bands and we would expect them to overlap with each other.
For our illustrative 2 oscillator model, this is what occurs.
For a large number of oscillators, the precise location of the overlap region is very difficult to estimate
if we have $q_i^{\mathrm{eff}} \gsim 1$.
Therefore, we will consider instead the limit when $q_i^{\mathrm{eff}} \ll 1$ for each of the oscillators
when the overlap occurs.
Since the most important resonances are the first order ones in this case, we will consider
where the first order resonances overlap.
For $q_i^{\mathrm{eff}} \ll 1$, the boundaries of the $i$th stability tongue are given by $A_{\mathrm{min}} = m_i^2/m^2 - q_i$
and $A_{\mathrm{max}} = m_i^2/m^2 + q_i$.~\cite{rrand1,rrand2}
Define $\delta m_{i,j}^2 = m_j^2 - m_i^2$.
Then the $i$th and $(i+1)$th first order bands overlap when
\begin{equation}
  \label{eqn:qoverlap}
  q_i + q_{i+1} \cong \frac{\delta m_{i,i+1}^2}{m^2} \, .
\end{equation}
For the purposes of illustration, suppose that the mass splittings and $q_i$'s are all equal
Denote $\delta m_{i,i+1}^2 = \delta \bar{m^2}$ and $q_i=\bar{q}$.
Then the above tells us that the first order resonance bands will overlap when
\begin{equation}
  \label{eqn:qoverlapeq}
  \bar{q} \sim \frac{\delta \bar{m^2}}{2m^2} \, .
\end{equation}
The effective $q$ value in the $i$th first order band is then 
\begin{equation}
  q_i^{\mathrm{eff},\mathrm{overlap}} \sim  \frac{\delta \bar{m^2}}{2m_i^2} \, .
\end{equation}
Hence, the above estimate is consistent provided the mass distribution is not too broad 
(\emph{i.e.} $\delta\bar{m^2} \ll m_i^2$ for each oscillator).

Finally, consider increasing $q$ even further, into the regime where the resonance bands are no longer in the process of overlapping.
Figure~\ref{fig:floquethugeq} demonstrates that the Floquet exponents become much simpler
functions of $A$ and the complicated effects of interference wash out.
Relative to the Mathieu equation, we see that the resonance is more efficient for essentially all choices of $A$. 
Additionally, there are fewer stable modes and that the resonance extends over a larger range of A values.

Before applying the above mathematical framework to the physical problems at hand, one final remark is in order.
Our above analysis of the Floquet exponents took a global viewpoint in that they were computed by determining the
growth of the mode functions over long time intervals.
However, during reheating, the resonance parameters evolve and only the early parts of the evolution of the mode functions tend to be relevant.
As well, when the resonance is strong, the mode functions will grow very quickly and our linear analysis soon ceases to be valid.
Again, it is only the early time evolution of the mode functions that tends to be relevant.
Thus, some knowledge of the form of the mode functions themselves (not just their average growth) is required in order to truly
apply the above formalism to reheating.
However, here we will be content primarily with deciding whether or not preheating is possible, in which case a knowledge
of the asymptotic Floquet exponents should be sufficient.

\EPSFIGURE[ht]{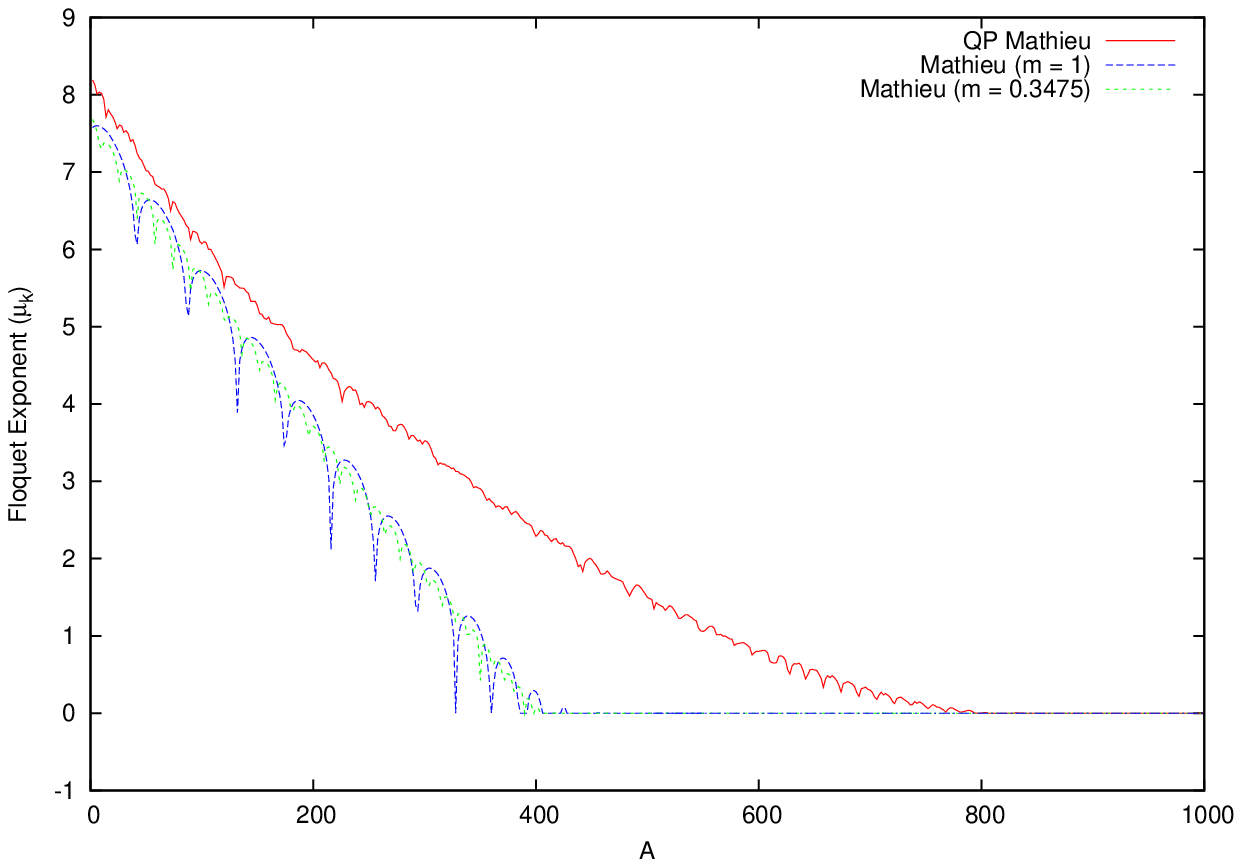,width=4in}{Floquet exponents for $q = 200$ for the QP Mathieu equation and the Mathieu equation with two different oscillation frequencies.  In the QP case, we see that far fewer modes are stable and that the resonance
remains strong for larger values of $A$.\label{fig:floquethugeq}}

\subsection{Thee-Legs Interaction: Geometrical Interpretation}

Having developed the mathematical theory of the QP Mathieu equation, let us now apply this insight to study reheating after N-flation.  We begin by considering the 3-legs interaction in the 
context discussed above.  Equation (\ref{3-thingy}) takes the form (\ref{hill_3}) with $2z = mt + \pi /2$ and the resonance parameters (ignoring terms involving $H/m$, which is the condition 
for $A$ and $q_i$ to evolve adiabatically)
\begin{equation}
\label{3legs_params}
  A_k = \frac{4 k^2}{m^2 a^2}, \hspace{5mm} q_i = \frac{2 \sigma_i \phi_{0,i}}{m^2 a^{3/2}} \, .
\end{equation}

Recall our picture of the parameters of our mode functions tracing out curves in the $(A,{\bf q})$ 
parameter space.  In this case, the initial parameter values are distributed along the vertical
line $q_{i,0} = \frac{2 \sigma_i \phi_{0,i}}{m^2 a_0^{3/2}}$ with $A \ge 0$.
Subsequently, each mode will evolve along the curve $A_k = A_{k,0}\left(\frac{q_i}{q_{i,0}}\right)^{4/3}$.
This evolution is illustrated in Fig.~\ref{fig:trace} for a particular choice of initial conditions $q_{i,0}$ 
(determined by $\phi_{0,i}$ and $\sigma_i$) and several $k$ values.  As can be seen from Fig.~\ref{fig:trace}, 
modes can pass through the broad instability regions where many of the resonance ``tongues'' have overlapped.
Roughly, the modes lying below $A = \sum_i 2q_i$ will undergo tachyonic resonance, which will dominate the preheating.
This matches our previous expectation from investigation of the effective mass (\ref{M3legs}).

\EPSFIGURE[ht]{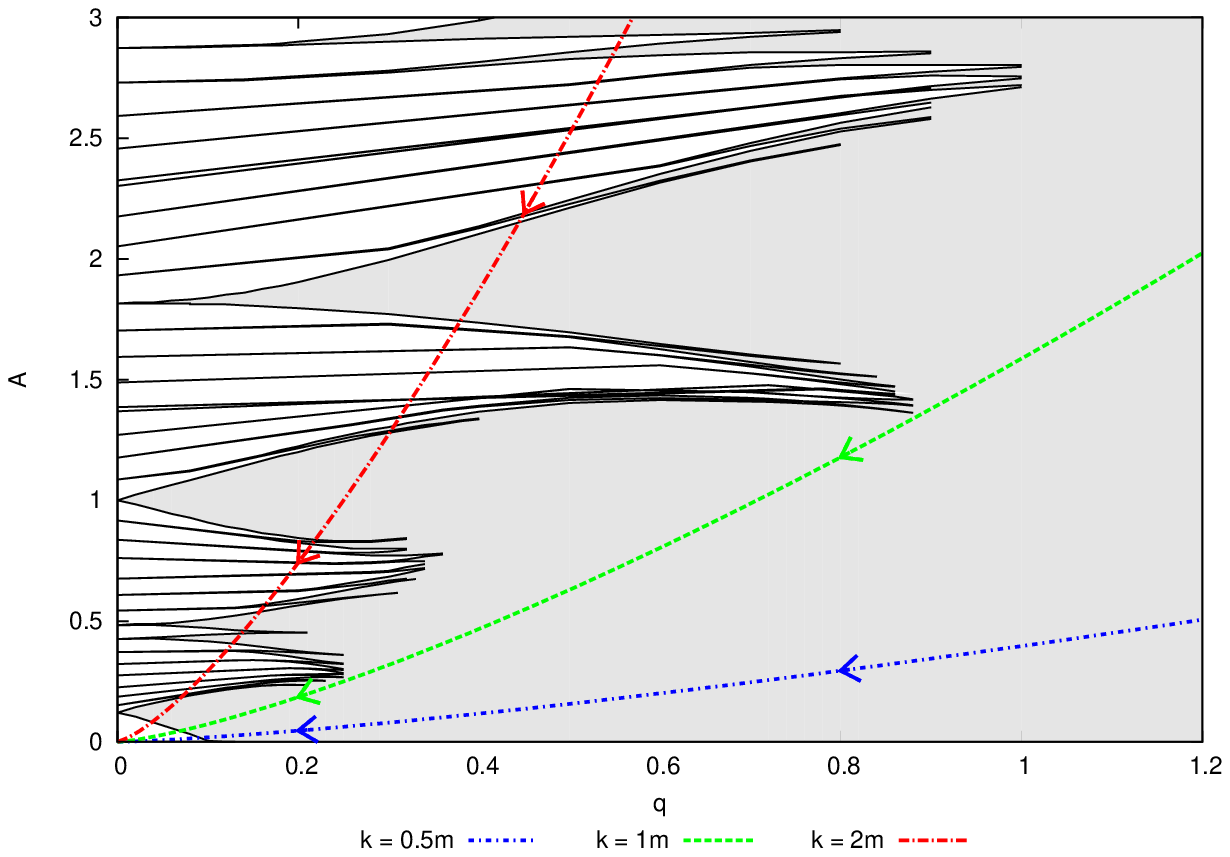,width=4in}{The paths through the instability chart traced out by several mode functions
(with different values of $k$) for the case of a 3-legs coupling between the inflatons and scalar field.
We have taken our initial value of $q$ to be $q_0 = 2$.
As time passes, all of the modes move towards the origin.\label{fig:trace}}

Our interpretation of excitations of the $\chi$ field in terms of particles allows us to give a physical
interpretation of the instability tongues in terms of Feynman diagrams (at least in the perturbative regime).
The instability regions correspond to those wave-numbers of the $\chi$ field that can be produced by the
oscillating inflaton condensate.
The only vertex available to us is the following:
\begin{center}
  \begin{picture}(244,112)(15,5)
    \SetWidth{1.0}
    \SetColor{Black}
    \ArrowLine(54,56)(144,56)
    \Text(54,56)[r]{$\phi_i$}
    \Vertex(144,56){1.5}
    \DashArrowLine(144,56)(234,16){1.0}
    \Text(234,16)[l]{$\chi$}
    \DashArrowLine(144,56)(234,96){1.0}
    \Text(234,96)[l]{$\chi$}
  \end{picture}
\end{center}
Since we have not quantized the inflaton, we have no inflaton propagators.
This allows for tree-level diagrams with exactly two outgoing $\chi$ lines.
These diagrams can be classified according to the net number of ingoing or outgoing lines
of each inflaton field.
To associate each of these classes of diagrams with a tongue, 
consider the tongue at $({\bf n} \cdot {\bf m})^2 = Am^2 = 4k^2a^{-2}$.
This value represents the energy squared available for the process.
To break the sign degeneracy, choose ${\bf n}$ such that ${\bf n} \cdot {\bf m} > 0$ is the energy available for the process.
Then $n_i$ represents the net number of ingoing (if $n_i > 0$) or outgoing (if $n_i < 0$)
inflaton source lines.
Since two $\chi$ particles are produced, energy-momentum conservation requires that they each
have $k_{\mathrm{phys}} = k/a = E/2$, thus giving $E^2 = ({\bf n}\cdot{\bf m})^2 = 4k^2a^{-2}$.
This is precisely the location of the tip of the resonance tongue labeled by ${\bf n}$.
As an example, the first diagram corresponding to the the $(2,-1)$ resonance for the case of
two fields are shown below (where we assume $2m_1 > m_2$):
\begin{center}
  \begin{picture}(244,112)(15,5)
    \SetWidth{1.0}
    \SetColor{Black}
    \ArrowLine(54,31)(144,31)
    \Text(54,31)[r]{$\phi_1$}
    \Vertex(144,31){1.5}
    \DashLine(144,31)(144,56){1.0}
    \Vertex(144,56){1.5}
    \ArrowLine(144,56)(214,56)
    \Text(216,56)[l]{$\phi_2$}
    \DashLine(144,56)(144,81){1.0}
    \ArrowLine(54,81)(144,81)
    \Text(54,81)[r]{$\phi_1$}
    \Vertex(144,81){1.5}
    \DashLine(144,56)(144,81){1.0}
    \DashArrowLine(144,31)(214,6){1.0}
    \Text(214,6)[l]{$\chi$}
    \DashArrowLine(144,81)(214,106){1.0}
    \Text(214,106)[l]{$\chi$}
  \end{picture}
\end{center}

We will develop the machinery to perform computations in the perturbative regime in the next section.
The reader should note that we are not claiming the perturbative series that can be extracted from our formalism
is identical to that familiar from Minkowski QFT.  Rather, this is intended as a useful picture to understand physically 
the dynamics involved.

\subsection{Four-Legs Interaction: Geometrical Interpretation}

Now, consider the situation for the 4-legs interaction.  Defining $z = mt$, the resonance parameters are
\begin{equation}
  A_k = \frac{k^2}{m^2 a^2} + 2\sum_i q_i, \hspace{5mm} q_i = \frac{g_i^2 \phi_{0,i}^2}{4m^2 a^3} \, .
\end{equation}
$A_k$ now has an explicit dependence on the $q_i$'s.
This results in the restriction
\begin{equation}
  2\sum_i q_i \le A
  \label{eqn:qrest}
\end{equation}
for the 4-legs interaction.  Compare this to $2q \le A$, which would be obtained in the single field case.
In Fig.~\ref{fig:trace2} where we plot the trajectory through parameter space which is induced due to the expansion
of the universe.
As in the above 3-legs case, we have an identification of the various instability tongues with various perturbative decay channels of the
inflaton.

\EPSFIGURE{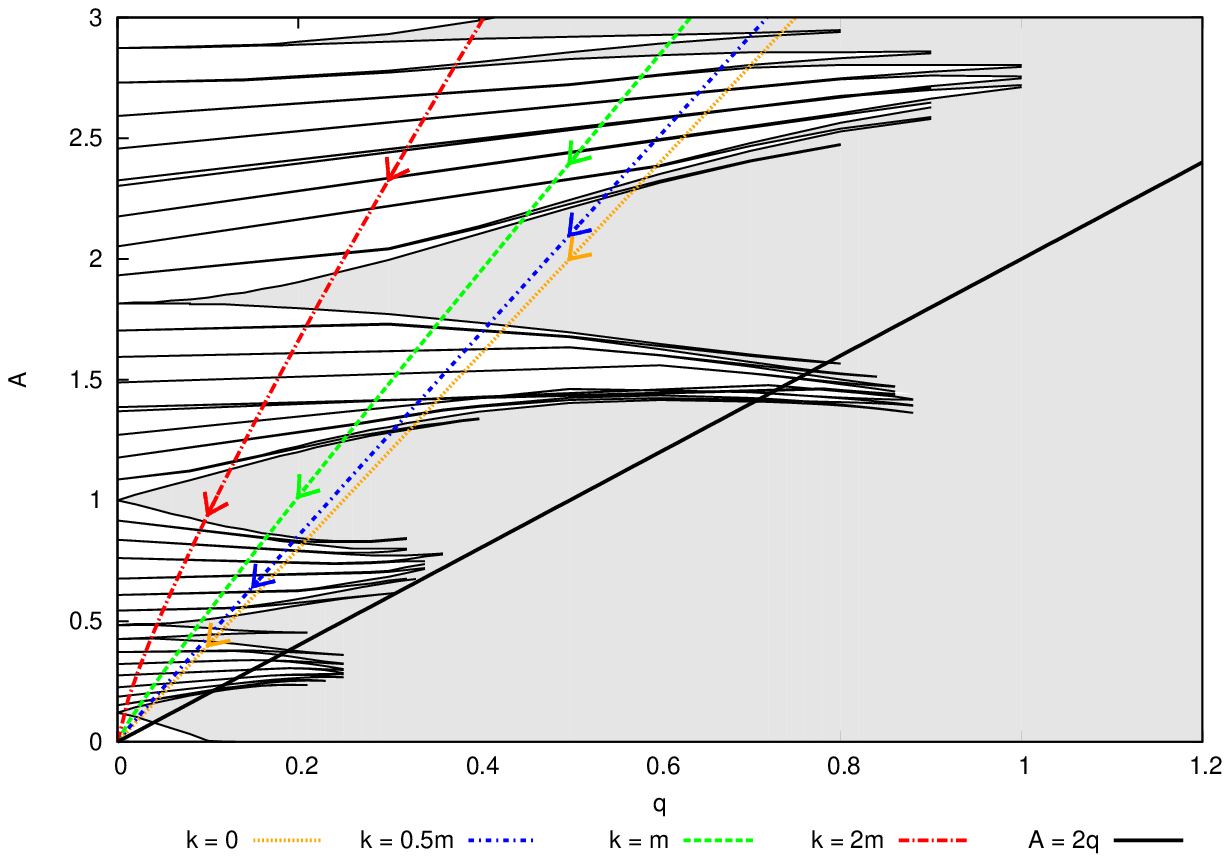,width=4in}{Paths followed by the resonance parameters for several mode functions for the case of 
a 4-legs interaction between the inflatons and scalar field.  Here we have chosen the parameter $q$ to take the
value $q_0 = 2$ at the start of reheating.  All of the modes lie above the path folled by the $k=0$ mode (orange dotted line).  This restriction prevents the modes from passing through the broad instability regions.
For illustrative purposes, the line $A=2q$ is also shown.
This is the line the modes must lie above in the single field case.\label{fig:trace2}}

Condition~\eqref{eqn:qrest} is ultimately the reason why nonperturbative decays of the inflaton are absent for the 4-legs interaction.
To see this, first suppose that the mass spectrum is such that the condition $2\sum_i q_i \le A$ ensures that the resonances are all distinct.\footnote{We'll address this point below.}
If we have a large number of $q_i$'s with similar values, then each one satisfies $2q_i \ll A$ for~\eqref{eqn:qrest} to hold.
Now consider the value of $q_i$ inside the $i$th first order resonance.
We have $2q_i\ll~ A~=~(m_i/m)^2$, which gives $q_i^{\mathrm{eff}}=q_i(m/m_i)^2\ll~1$.
Therefore, \emph{if} the resonance bands are distinguishable, we have $q_i^{\mathrm{eff}} \ll 1$ and the decays of the inflatons will be perturbative.

The above argument relies on a notion of distinct resonance bands for each oscillator.
Let's again assume that all of our squared mass splittings and $q_i$'s are equal.
From~\eqref{eqn:qrest} and~\eqref{eqn:qoverlap} with $A=(m_i/m)^2$ in the $i$th resonance band,
we see that for \emph{none} of the resonance bands to have overlapped we require
\begin{equation}
  \label{eqn:massrest}
  N_{\mathrm{osc}}\delta\bar{m^2} \gsim m_i^2
\end{equation}
for all masses $m_i$.
Actually, the above estimate is quite conservative.
Since $q_i^{\mathrm{eff}} \ll 1$ for each individual oscillator, in order to have a strong resonance we must have many
different oscillators contributing.
Thus, rather than having only pairs of instability bands overlapping, we must have a large number of them overlapping simultaneously,
which will require an even more degenerate mass spectrum than the above estimate.
In agreement with our investigation of the dynamics of the effective mass (\ref{M4legs}), 
we conclude that provided the mass spectrum satisfies~\eqref{eqn:massrest} the decays via the 4-legs coupling will always
be perturbative.
In this respect we differ from the conclusions of~\cite{cantor1,cantor2}, where the addition of an additional
oscillating scalar was predicted to increase the efficiency of the resonance.
As well, for the case of many oscillating fields, the resulting dynamics should be similar to the case where the
driving term is random~\cite{noise1,noise2}.
However, even for the case of many fields, we find here that for small $q$ the resonance effects will be too weak to lead to relevant instabilities for all but a small subset of modes.
This will be confirmed also in section~\ref{sec:4legs} below.

To reinforce the point above, in Fig.~\ref{fig:floq_quad} we show the Floquet exponents along the lines $A=2q$ and $A=4q$ both for the two oscillator case above and  
the Mathieu equations of the individual oscillators.  Although the presence of the second field strengthens the resonance along $A=2q$, it simultaneously forces the 
modes above $A=4q$ actually decreasing the strength of the resonance as seen by the mode functions.
However, preheating is still possible for the case of two oscillators, although it is suppressed relative
to the single field case~\cite{multi_preheat2}.


\EPSFIGURE{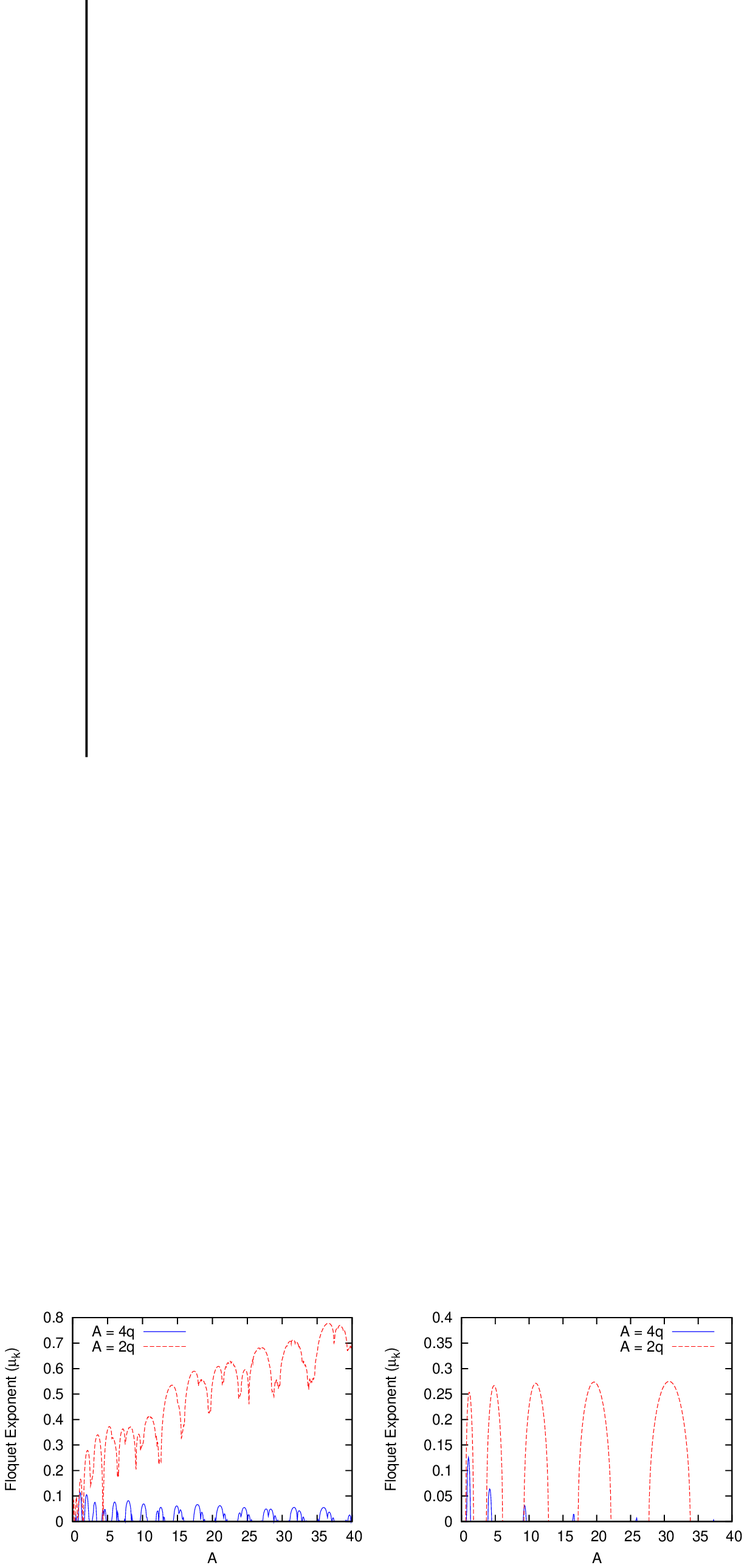}{Floquet exponents along the curves $A=4q$ (blue curve) and $A=2q$ (red curve) 
for both the QP Mathieu (left panel) and periodic Mathieu (right panel) equations.
The resonance is clearly much stronger along the line $A=2q$ in both cases.
Additionally, the resonance is stronger in the QP case along both lines (note the scales on the
vertical axes).\label{fig:floq_quad}}

\subsection{On the Inclusion of Off-Diagonal Interactions}


In our analysis thus far, we have neglected the possibility of off-diagonal 4-leg interactions of the form
\begin{equation}
\label{off-diag}
\mathcal{L}_{\mathrm{int}}^{\mathrm{off-diag}}= \sum_{i\not= j}\frac{h_{ij}}{2}\phi_i\phi_j \chi^2
\end{equation}
in the prototype action (\ref{L2}).  However, if $\sigma_i\not= 0$ then no symmetry forbids such terms and, in general, they must be included.  Let us now argue that their consistent inclusion does not add anything qualitatively
new to our previous discussion.  Note that the interaction (\ref{off-diag}) will yield the following contribution to the effective mass for the $\chi$ fluctuations:
\begin{equation}
  \Delta M_{\mathrm{eff}}^2(t) = \sum_{i\not= j} h_{ij} \phi_i(t)\phi_j(t) \, .
\end{equation}
Using the solution (\ref{Nosc}) we have
\begin{eqnarray}
  \Delta M_{\mathrm{eff}}^2(t) &=&  \sum_{i\not= j} h_{ij} \frac{\phi_{0,i}\phi_{0,j}}{a^3(t)} \sin\left[m_i t + \theta_i\right]\sin\left[m_jt+\theta_j\right] \nonumber \\
  &=&  \sum_{i\not= j} h_{ij} \frac{\phi_{0,i}\phi_{0,j}}{2 a^3(t)} \left[  \sin\left(m_{ij}^{-} t + \theta_{ij}^{-}\right) -  \sin\left(m_{ij}^{+} t + \theta_{ij}^{+}\right) \right] \, ,\label{off-diag-mass}
\end{eqnarray}
where on the last line we have used some elementary trigonometric identities and defined $m_{ij}^{\pm} = m_i \pm m_j$, $\theta_{ij}^{\pm} = \theta_i \pm \theta_j + \pi / 2$.  
Equation (\ref{off-diag-mass}) has precisely the same form as the contribution from 3-leg interactions of the form $\sum_i\sigma_i \phi_i\chi^2$.  The only difference is the factor $a^{-3}$, 
which would have been $a^{-3/2}$ for trilinear couplings.  This difference does not play an important role for the short time scales relevant for nonperturbative preheating.
We conclude that off-diagonal 4-leg interactions (\ref{off-diag}) may lead to nonperturbative preheating by tachyonic resonance.

\section{Perturbative Decays of a Single Oscillating Inflaton}
\label{sec:pert}

In the last section we studied nonperturbative preheating effects at the end of N-flation.  In the case of 4-leg interactions we showed that strong nonperturbative effects are absent.  Hence, the decay of the N-flatons in this case
must proceed via perturbative preheating decays.  On the other hand, in the case of 3-leg interactions we found that strong nonperturbative effects are possible via tachyonic resonance.  However, perturbative processes may still be relevant
in this case also, either during the final stages of reheating or when the amplitude of the inflaton oscillations is small.  Thus, we now turn our attention to studying the perturbative decays of the homogeneous oscillating inflatons after
N-flation.  

Before we consider the case at hand, $N\gg 1$, it is interesting and instructive to first treat the simpler case of single field inflation, $N=1$, as a warm-up exercise. 
The material in this section 
is partially review~\cite{ref:kofman96,ref:KLSpert} and hence the reader who is already familiar with the theory of reheating may wish to skip ahead to the next section.  Here we explain the failure of the inflaton condensate to decay
via 4-leg interactions and also show that 3-leg interactions \emph{do} allow for a complete decay.  Later, we will show that the same results persist in the interesting case with $N\gg 1$.

As in our multi-field analysis, the homogeneous oscillations of the inflaton are described by 
\begin{equation}
\label{1osc}
  \phi(t) \cong \frac{\phi_0}{a^{3/2}(t)} \sin( m t + \theta ) \, .
\end{equation}
The universe expands as $a(t) \sim t^{2/3}$ and the energy density in $\phi$ red-shifts like non-relativistic dust: $\rho_\phi = \frac{1}{2}\dot{\phi}^2 + \frac{m^2}{2}\phi^2 \sim a^{-3}$.  As in (\ref{L2}) 
we couple the inflaton to a single scalar field via both 4-leg and 3-leg interactions as
\begin{equation}
\label{L1}
  \mathcal{L} = -\frac{1}{2}\partial_\mu\phi\partial^\mu\phi - \frac{m^2}{2} \phi^2 - \frac{1}{2}\partial_\mu\chi\partial^\mu\chi 
  - \left[ \frac{g^2}{2}\phi^2 + \frac{\sigma}{2}\phi  \right]\chi^2 - \frac{\lambda}{4} \chi^4 \, .
\end{equation}
The term $\lambda\chi^4$ is required to keep the potential bounded from below, but otherwise will play no role in our discussion.
We restrict to $\lambda \geq \sigma^2 / (2m^2)$ so that the potential has a single minimum $\phi=\chi=0$ with $V=0$.  
From (\ref{L1}) we can derive the equation of motion for the linear inhomogeneous fluctuations of the $\chi$ field in the background of a classical, homogeneous inflaton condensate $\phi$:
\begin{equation}
\label{chi_eqn}
  \ddot{\chi} + 3 H \dot{\chi} - \frac{\nabla^2}{a^2} \chi + \left[ g^2\phi^2(t) + \sigma \phi(t)   \right]\chi = 0 \, .
\end{equation}

Notice that the combination of 3-leg ($\phi\chi^2$) and 4-leg ($\phi^2\chi^2$) 
interactions in (\ref{L1}) may arise in a number of ways.  For example, if the inflaton potential $V(\tilde{\phi})$ has a minimum at $\tilde{\phi} = v$ about which $V(\tilde{\phi}) \cong m^2 (\tilde{\phi}-v)^2 / 2$ then, 
after the shift $\tilde{\phi} = \phi + v$, we have the familiar form $V \cong m^2\phi^2/2$.  A bi-linear coupling such as $g^2 \tilde{\phi}^2 \chi^2$ then leads 
to the addition of a 3-legs interaction $\sigma \phi \chi^2$ (with $\sigma = 2 g^2 v$)
after such a spontaneous symmetry breaking.  The scalar potential in (\ref{L1}) may also arise from the super-potential 
\begin{equation}
  W = \frac{m}{2\sqrt{2}} \phi^2 + \frac{g}{2\sqrt{2}} \phi\chi^2 \, ,
\end{equation}
which gives $\lambda = g^2/2$ and $\sigma = g m$.  In our perturbative analysis below it will not be necessary to assume that either the 3-leg or 4-leg interaction dominates, we can treat both simultaneously.

\subsection{Bogoliubov Calculation}
\label{subsec:pert}

We now study the perturbative production of $\chi$ particles from the decay of the oscillating field (\ref{1osc}).  
We treat the inflaton field as a classical background and study the linear quantum fluctuations of the inhomogeneous field $\chi(t,{\bf x})$.
It is convenient to work with conformal time $\tau$, defined in terms of the usual cosmic time variable $t$ as $a d\tau = dt$, and
introduce a re-scaled ``co-moving'' field 
\begin{equation}
  f(\tau,{\bf x}) = a(\tau) \chi(\tau,{\bf x}) \, .
\end{equation}
We decompose the co-moving field into q-number 
annihilation/creation operators $a_{\bf k}$, $a_{\bf k}^{\dagger}$ and c-number mode functions $f_k(\tau)$ as
\begin{equation}
  f(\tau,{\bf x}) = \int \frac{d^3 k}{(2\pi)^{3/2}} \left[ a_{\bf k} f_k(\tau) e^{i {\bf k}\cdot {\bf x}} + a_{\bf k}^\dagger f_k^\star(\tau) e^{-i {\bf k}\cdot {\bf x}}\right] \, .
\end{equation}
The annihilation/creation operators obey the usual commutation relation
\begin{equation}
  \left[ a_{\bf k}, a_{\bf k'}\right]= \delta^{(3)}({\bf k} - {\bf k'})
\end{equation}
 and the mode functions satisfy the oscillator-like equation
\begin{equation}
\label{oscillator}
  f''_{k}(\tau) + \omega_k^2(\tau) f_k(\tau) = 0 \, ,
\end{equation}
where the prime denotes a derivative with respect to cosmic time $f' \equiv \partial_\tau f$ and the time-varying frequency is
\begin{eqnarray}
  \omega_k^2(\tau) &=& k^2 + a^2 \left[ g^2 \phi^2(\tau) + \sigma \phi(\tau) \right] - \frac{a''}{a} \nonumber \\
                   &\cong& k^2 + \frac{g^2 \phi_0^2}{a(\tau)} \sin^2\left[m t(\tau) + \theta\right] + a^{1/2}(\tau)\sigma \phi_0 \sin\left[m t(\tau) + \theta\right] \, , \label{freq}
\end{eqnarray}
where $t(\tau) = \int a d\tau$.  On the second line of (\ref{freq}) we have used (\ref{1osc}) 
and the fact that $a'' / a \sim a^2 H^2$ is negligible compared to the modes $k^2 \sim a^2 m^2$ that we expect to be produced.  (Recall that $H \ll m$ during reheating.)

We seek a solutions of equation (\ref{oscillator}) of the WKB form
\begin{align}
  f_k(\tau) &= \alpha_k(\tau) \frac{e^{-i \int \omega_k(\tau') d\tau'}}{\sqrt{2\omega_k(\tau)}} + \beta_k(\tau) \frac{e^{+i \int \omega_k(\tau') d\tau'}}{\sqrt{2\omega_k(\tau)}} \, ,\label{bog} \\ 
  f_k'(\tau) &= -i\alpha_k(\tau) \sqrt{\frac{\omega_k(\tau)}{2}}e^{-i \int \omega_k(\tau') d\tau'} + i\beta_k(\tau)\sqrt{\frac{\omega_k(\tau)}{2}}e^{+i \int \omega_k(\tau') d\tau'} \, ,\label{bog2}
\end{align}
where the Bogoliubov coefficients are normalized as $|\alpha_k|^2 - |\beta_k|^2 = 1$.  The ansatz (\ref{bog},\ref{bog2}) affords a solution of (\ref{oscillator}) provided $\alpha_k(\tau)$, $\beta_k(\tau)$ obey the coupled equations
\begin{equation}
\label{bog_ev}
  \alpha'_k(\tau) = \frac{\omega'_k(\tau)}{2\omega_k(\tau)} e^{+2 i \int \omega_k(\tau') d\tau'} \beta_k(\tau), \hspace{5mm} 
  \beta'_k(\tau) = \frac{\omega'_k(\tau)}{2\omega_k(\tau)} e^{-2 i \int \omega_k(\tau') d\tau'} \alpha_k(\tau) \, .
\end{equation}
The occupation number $n_k$ is defined by the energy of the mode $\frac{1}{2}|f'_k|^2 + \frac{1}{2}\omega_k^2 |f_k|^2$ divided by the frequency of that mode
\begin{eqnarray}
  n_k &=& \frac{1}{2\omega_k} \left[ |f'_k|^2 + \omega_k^2 |f_k|^2  \right] - \frac{1}{2} \nonumber \\
      &=& |\beta_k|^2 \label{n_k} \, 
\end{eqnarray}
where the $-\frac{1}{2}$ comes from extracting the zero point energy of the harmonic oscillator and on the second line we have assumed the WKB solution (\ref{bog}).  The occupation number depends only
on the negative frequency Bogoliubov coefficient, $\beta_k$.

Demanding that the field be in the adiabatic vacuum in the asymptotic past gives the initial condition $\alpha_k = 1$, $\beta_k = 0$. In the perturbative regime particle occupation numbers remain small $|\beta_k| \ll 1$
so we can iterate (\ref{bog_ev}) to obtain
\begin{equation}
\label{beta_k}
  \beta_k(\tau) \cong \int_{-\infty}^{\tau}d\tau' \frac{\omega'_k(\tau')}{2 \omega_k(\tau')} \exp\left[ -2 i \int_{-\infty}^{\tau'} d\tau'' \omega_k(\tau'')  \right] \, .
\end{equation}
To evaluate the integral (\ref{beta_k}) we consider the limit
\begin{equation}
\label{limits_stuff}
   \frac{k^2}{a^2} \gg g^2 \phi^2, \hspace{5mm} \frac{k^2}{a^2} \gg \sigma \phi \, .
\end{equation}
For the modes  $k \sim a m$ that we expect to be produced, these limits are equivalent to assuming that the resonance parameters of the Mathieu equation are small: 
$q_3 \equiv \sigma \phi_0 / m \ll 1$ and $q_4 \equiv g^2 \phi_0^2 / m^2 \ll 1$.  Hence, the limits (\ref{limits_stuff}) are compatible with the assumption that nonperturbative
preheating processes are negligible.

Given (\ref{limits_stuff})  we can approximate $e^{-i \int \omega_k d\tau'} \cong e^{-i \int k d\tau'} \cong e^{-i k \tau}$.  On the other hand, for the frequency (\ref{freq}) we have
\begin{eqnarray}
  \frac{\omega'_k}{\omega_k} &\cong& \frac{a^2}{2\omega_k^2} \left[ 2 g^2  \phi + \sigma  \right] \frac{d\phi}{d\tau} \nonumber \\
  &\cong& \frac{a^{3/2}\phi_0 m}{2\omega_k^2}\left[ \frac{2 g^2 \phi_0}{a^{3/2}} \sin\left[m t(\tau)+\theta\right]  + \sigma \right]
    \cos\left[m t(\tau) + \theta\right] \, ,
\end{eqnarray}
where we neglected terms with derivatives of the scale factor, $a$.  Collecting everything together, we have
\begin{eqnarray}
\beta_k(\eta) &\cong& \frac{g^2\phi_0^2m}{8ik^2} \int_{-\infty}^{\tau} d \tau'\left(e^{ik \psi_{4-\mathrm{leg}}^{+}(\tau')}-e^{-ik \psi_{\mathrm{4-leg}}^{-}(\tau')}\right) \nonumber \\
  && + \frac{\sigma\phi_0m}{8k^2}
\int_{-\infty}^{\tau} d\tau' a^{3/2} \left(e^{ik \psi_{3-\mathrm{leg}}^{+}(\tau')}-e^{-ik \psi_{3-\mathrm{leg}}^{-}(\tau')}\right) \, ,  \label{eqn:singlefield}
\end{eqnarray}
where we have defined
\begin{equation}
  \psi_{4-\mathrm{leg}}^{\pm}(\tau) = -2\tau \pm \frac{2mt(\tau)+2\theta}{k} \qquad 
\mathrm{and}
\qquad \psi_{3-\mathrm{leg}}^{\pm}(\tau) = -2\tau \pm \frac{mt(\tau)+\theta}{k} \ .
\end{equation}
The first term in (\ref{eqn:singlefield}) is the contribution from the 4-leg interaction $\phi\phi\rightarrow \chi\chi$ while the second term is the contribution from the 3-leg interaction $\phi\rightarrow\chi\chi$.
Notice that the integrals appearing in (\ref{eqn:singlefield}) are defined by the interference of two oscillatory terms, $e^{imt(\tau')+\theta}$ and $e^{-2ik\tau'}$ 
or $e^{2imt(\tau')+2\theta}$ and $e^{-2ik\tau'}$,  where the 
function $t(\tau) = \int a d\tau$ is non-linear.  
We will give a physical interpretation of these factors later.

The stationary phase approximation tells us that the integrals appearing in (\ref{eqn:singlefield}) are dominated by the contribution near the instant where
\begin{equation}\label{inst}
\frac{d}{d\tau}  \psi_{n-\mathrm{leg}}^{+}(\tau) = 0 \ .
\end{equation}
For the 4-leg interaction, this is the moment $\tau = \tau_{4,k}$ when  
\begin{equation}\label{four}
ma(\tau_{4,k})=k \ ,
\end{equation}
while for the 3-leg interaction the dominant contribution to (\ref{eqn:singlefield}) comes from the moment $\tau=\tau_{3,k}$ when
\begin{equation}\label{tri}
ma(\tau_{3,k})=2 k \ .
\end{equation}
These results are simple to understand on physical grounds: they reflect energy conservation for an annihilation of two
inflaton particles at rest into a pair of $\chi$'s, and decay of a single inflaton into a pair of $\chi$'s, respectively.

Now we can calculate the co-moving occupation number (\ref{n_k}):
\begin{eqnarray}
  n_k &=& |\beta_k|^2  \label{eqn:numkonefield} \\
      &=& \frac{\pi g^4 \phi_0^4 m^3 }{64 k^6 H_{4,k}} \Theta\left[a(t)-a_{4,k}\right]\Theta\left[a_{4,k}-a_0\right]   + \frac{\pi \sigma^2 \phi_0^2}{16 k^3 H_{3,k}} \Theta\left[a(t)-a_{3,k}\right]\Theta\left[a_{3,k}-a_0\right] \nonumber \\
      && + \frac{\pi g^2 \sigma \phi_0^3 m^{3/2}}{16 k^{9/2} (H_{4,k}H_{3,k})^{1/2}} \sin \left[ \psi_{4-\mathrm{leg}}^{+}(t_{4,k}) -  \psi_{3-\mathrm{leg}}^{+}(t_{3,k})  \right]
         \Theta\left[a(t)-a_{3,k}\right]\Theta\left[a_{4,k}-a(t)\right] \, . \nonumber
\end{eqnarray}
The formula (\ref{eqn:numkonefield}) is the main result of this subsection.  The subscript $_{4,k}$ indicates that the quantity is to be evaluated at the 
moment $\tau=\tau_{4,k}$, when $k=ma$ and the 4-leg interaction proceeds for a given $k$.
Similarly,
the subscript $_{3,k}$ indicates that a quantity must be evaluated at $\tau=\tau_{3,k}$ when $2k = ma$ and the 3-leg interaction proceeds for a given $k$.  
We denote the scale factor at the beginning of reheating by $a_0$.  The step functions $\Theta$ 
appearing in (\ref{eqn:numkonefield}) enforce the fact that in order for the $\chi$'s to be produced perturbatively in pair annihilation and decays of inflatons by time $t$, energy conservation requires that $ma_0 < k < ma(t)$ 
and $ma_0 < 2k < ma(t)$ respectively.

\subsection{Boltzmann Equation}
\label{subsec:boltz}

We define the co-moving energy density in the $\chi$ field as 
\begin{align}
a^4\rho_{\chi} &\equiv \int \frac{d^3k}{(2\pi)^{3}}\,  \omega_k n_k \nonumber \\ \nonumber
&= \frac{g^4\phi_0^4}{128 \pi}\int_{ma_0}^{ma(t)}d k\frac{m^3}{H_{4,k}k^3}
+\frac{\sigma^2\phi_0^2}{32\pi}\int_{ma_0/2}^{ma(t)/2}d k\frac{1}{H_{3,k}}\\
&\qquad +\frac{g^2\sigma\phi_0^3}{32\pi}\int_{ma_0}^{ma(t)/2}d k \frac{m^{3/2}}{k^{3/2}\sqrt{H_{4,k}H_{3,k}}}
\sin\left[\psi_{4-\mathrm{leg}}^{+}(t_{4,k})-\psi_{3-\mathrm{leg}}^{+}(t_{3,k}) \right].
\end{align}
The time rate of change is then given by
\begin{eqnarray}
a^{-4}\frac{\mathrm{d}}{\mathrm{d}t}(a^4\rho_{\chi})
&=& \frac{g^4\phi_0^4m}{128\pi a^6}
+ \frac{\sigma^2\phi_0^2m}{64\pi a^3} \nonumber \\
&+& \frac{\sqrt{2}g^2\sigma\phi_0^3}{32\pi a^{9/2}} \sqrt{\frac{H(a)}{H(a/2)}}\sin\left[\psi_{4-\mathrm{leg}}^{+}(\tau(a/2))-\psi_{3-\mathrm{leg}}^{+}(\tau(a))\right] \, .\label{boltz1}
\end{eqnarray}
The first two terms are the familiar result for the Boltzmann equation with $\phi\phi\rightarrow\chi\chi$ and $\phi\rightarrow\chi\chi$ processes contributing to the collision integral.
The final term in this expression arises due to quantum mechanical interference between the two decay channels of the inflaton.  
It is present because the inflaton is a condensate (which we treat as a classical source for the quantum $\chi$ field) and not a collection of particles.

Notice that the oscillatory ``interference'' term in (\ref{boltz1}) disappears when we average over time and we are left with the usual Boltzmann equation
\begin{equation}
\label{eqn:oneinfenergy}
a^{-4}\frac{d}{\mathrm{d}t}(a^4\rho_{\chi})
\cong 2\frac{[\sigma_{\phi\phi\rightarrow\chi\chi} v ]_{v=0}}{m}\rho_{\phi}^2
+ \Gamma_{\phi\rightarrow\chi\chi}\rho_{\phi} \, ,
\end{equation}
where we have used the fact that the energy density in the inflaton oscillations is 
$\rho_\phi = \frac{1}{2}(\dot{\phi}^2 + m^2 \phi^2) \cong \frac{m^2 \phi_0^2}{2 a^3}$.
The factor of $2$ in front of the annihilation cross-section in (\ref{eqn:oneinfenergy}) 
arises because there are two inflatons annihilating in each interaction.

By comparing (\ref{eqn:oneinfenergy}) to (\ref{boltz1})  we can identify the annihilation cross-section at zero relative velocity
\begin{equation}
\label{cross_sec}
[\sigma_{\phi\phi\rightarrow\chi\chi} v]_{v=0} = \frac{g^4}{64\pi m^2}
\end{equation}
and decay width
\begin{equation}
\label{rate}
\Gamma_{\phi\rightarrow\chi\chi} = \frac{\sigma^2}{32\pi m} \, .
\end{equation}
The rate (\ref{rate}) agrees with the standard result obtained in perturbative QFT for the decay of inflaton particles via $\phi\rightarrow \chi\chi$.  
The cross-section (\ref{cross_sec}) also agrees with the perturbative QFT result for inflaton particles at rest, provided that the Feynman amplitude is 
evaluated at $v=0$, so that the two incoming inflatons are treated as identical particles in the combinatorial counting.  (Treating the inflatons as a 
classical source term accounts for this automatically.) 

As usual, conservation of energy-momentum requires $a^{-3}\frac{\ud}{\ud t}(a^3\rho_{\phi}) = -a^{-4}\frac{\mathrm{d}}{\mathrm{d}t}(a^4\rho_{\chi})$.
This allows us to write~\eqref{eqn:oneinfenergy} as
\begin{equation}
\label{comoving_decay}
\frac{d}{d t}(a^3\rho_{\phi}) = -2\frac{[\sigma_{\phi\phi\rightarrow\chi\chi} v]_{v=0}}{ma^3}(a^3\rho_{\phi})^2
- \Gamma_{\phi\rightarrow\chi\chi}(a^3\rho_{\phi}) \, .
\end{equation}
This equation has a simple physical interpretation.  In the absence of any interactions, it shows that the co-moving energy density of inflaton particles 
remains constant: $a^3 \rho_\phi \sim \mathrm{const}$.  On the other hand, if we include only the decays $\phi\rightarrow\chi\chi$ then equation (\ref{comoving_decay}) shows
that the co-moving energy density must decrease exponentially: $a^3 \rho_\phi \sim e^{-\Gamma t}$.  In a time of order $\Gamma^{-1}$ the inflaton condensate has decayed completely.

It is interesting to consider equation (\ref{comoving_decay}) in the special case where annihilations $\phi\phi\rightarrow\chi\chi$ are present but decays $\phi\rightarrow\chi\chi$ are excluded.
In this case we can integrate (\ref{comoving_decay}) to show that $a^3 \rho_\phi \rightarrow \mathrm{const}$ as $t \rightarrow \infty$.  In other words, the inflaton does not completely decay and
some finite co-moving density of non-relativistic $\phi$ ``particles'' freezes out. The problem is that the volume dilution of the non-relativistic particles proceeds faster than the annihilation 
process $\phi\phi\rightarrow\chi\chi$ can drain energy from the condensate.  Typically, this failure to decay will leave a late-time universe which is cold and unsuitable for life, not unlike 
certain parts of Canada.  Clearly this is an unacceptable scenario.  Our analytical conclusion that 4-leg interactions do not allow the inflaton to decay is consistent
with the results of numerical lattice field  theory simulations, see \cite{eqn_state} for example.

\subsection{Comparison to Inflaton Particle Decays}

In this section we have studied the decay of the coherent, classical oscillations of the inflaton field (\ref{1osc}) into quanta
of the $\chi$ field.  It is instructive to compare our results to what one would have obtained for inflaton \emph{particles} 
with the couplings (\ref{L1}) to matter.  The key discrepancy is the appearance of the ``interference'' term on the last line of (\ref{boltz1}), which averages 
to zero.  As stated above, this term represents quantum interference between the decay channels of the inflaton and it arises because we have 
a coherent inflaton condensate (allowing us to treat $\phi$ as a classical source).

To be clear: in our formalism we are describing the production of pairs of $\chi$ particles from a coherent classical inflaton 
\emph{condensate}, as opposed to the production of $\chi$'s from annihilations/decays of inflaton \emph{particles}.  These processes
are essentially different, even though the final decay products are the same.   For the decay of classical inflaton oscillations, the amplitudes for the 3-leg and 4-leg interactions must be added coherently.
This results in an interference term when the amplitude is squared -- the sinusoidal term on the last line of (\ref{boltz1}).

\section{Perturbative Inflaton Decays via Four-Leg Interactions}
\label{sec:4legs}

\subsection{Bogoliubov Calculation}

In section~\ref{sec:preheating} we excluded nonperturbative effects for the 4-legs interaction $\sum_i g_i^2 \phi_i^2\chi^2$.
 We now study perturbative decays of the inflatons via this coupling.
Although we have the specific example of N-flation in mind, the methods in this and the following section should apply to
the late time perturbative decay of the inflaton condensate in any multi-field model where the condensate oscillates around
a minimum in its potential with small amplitude at late times.
Our approach follows very closely the formalism developed in subsections \ref{subsec:pert} and \ref{subsec:boltz}.  
The reader is encouraged to review these sections for definitions and more details of the derivation.  The quantum eigenmodes
of the $\chi(t,{\bf x})$ field obey the oscillator-like equation (\ref{oscillator}) with effective frequency
\begin{eqnarray}
  \omega_k^2(\tau) &=& k^2 + a^2 \sum_i g_i^2 \phi_i^2(\tau) - \frac{a''}{a} \nonumber \\
                               &\cong& k^2 + \frac{1}{a(\tau)}\sum_i g_i^2 \phi_{0,i}^2 \sin^2\left[m_i t(\tau) + \theta_i \right]  \, .\label{omega4legs}
\end{eqnarray}
On the last line we have used (\ref{Nosc}) and again neglected $a'' / a \sim a^2 H^2$.
The WKB form (\ref{bog}) still holds with $\alpha_k \cong 1$ and $\beta_k$ given by the expression (\ref{beta_k}).  

We wish to evaluate the integrals in (\ref{beta_k}) for the relevant effective frequency (\ref{omega4legs}).  To this end, we can consider the short wavelength limit
\begin{equation}
\label{eqn:klimit}
k^2 \gg a^2 \sum_{i}g_i^2\phi_{i}^2 \ .
\end{equation}
To see that this is the relevant assumption, note that modes with $k \sim m_ia$ are expected  to be produced. Therefore
the condition (\ref{eqn:klimit})  corresponds to the smallness of the effective resonance parameter in the Mathieu equation picture: $\sum_i q_i \ll 1$;
see subsection \ref{subsec:QP}.  Hence,  limit (\ref{eqn:klimit}) is compatible with our previous claim that strong nonperturbative effects are 
absent.

Using our background inflaton solutions \eqref{Nosc} and keeping only the leading order terms in $H/m_i$, (\ref{beta_k}) becomes
\begin{equation}
\beta_k(\tau) \cong \frac{1}{4k^2}\sum_iJ_{k}^{(i)}(\tau) \ ,
\end{equation}
where we have defined
\begin{equation}
J_{k}^{(i)}(\tau)=-\frac{g_i^2\phi_{0,i}^2 m_i}{2i}\int_{-\infty}^{\tau}d\tau'\left[e^{ik\psi_i^{+}(\tau')}-e^{-ik\psi_{i}^{-}(\tau')}\right] \ ,
\end{equation}
\begin{equation}
\label{fphase}
\psi_i^\pm(\tau)=-2\tau \pm \frac{2m_it(\tau)+2\theta_i}{k} \ .
\end{equation}
The integral is dominated by the contribution near the instant  $\tau=\tau_{i,k}$ when 
\begin{equation}
\label{eqn:annihtime}
m_i a(\tau_{i,k})= k \ .
\end{equation}
As before, \eqref{eqn:annihtime} has a simple interpretation: it reflects energy conservation for the annihilation of a pair of $\phi_i$ at rest
into a pair of massless $\chi$ particles with physical momentum ${\bf k} / a$. 

We can organize the inflatons so that $m_i < m_j$ for $i < j$.
Upon evaluating $J_k^{(i)}$, the expectation value of the co-moving occupation number of the $\chi$ particles is given by
\begin{align}
\label{eqn:numk}
n_{\bf k}=|\beta_k|^2 \approx &\frac{\pi}{64 k^3}
\sum_i\bigg\{\frac{g_i^4\phi_{0,i}^4}{a_{i,k}^3H_{i,k}}\Theta(t_{i,k}-t_0)\\
&+2\sum_{j>i}\frac{g_i^2g_j^2\phi_{0,i}^2\phi_{0,j}^2}{\sqrt{a_{i,k}^3H_{i,k}a_{j,k}^3H_{j,k}}}
\cos(k[\psi^{+}(t_{i,k})-\psi^{+}(t_{j,k})]) \, \Theta(t_{j,k}-t_0)
\bigg\} \Theta(t-t_{i,k}) \, , \nonumber
\end{align}
with $\psi^{+}_i(\tau)$ given by \eqref{fphase}, $t_{i,k}$ the cosmic time corresponding to conformal time $\tau_{i,k}$ and $t_0$ corresponds to 
the onset of reheating. 
We illustrate the accuracy of this approximation in Figure~\ref{fig:4legpertgraph}.
This is an accurate approximation provided $|\beta_k| \ll 1$. 

\EPSFIGURE[ht]{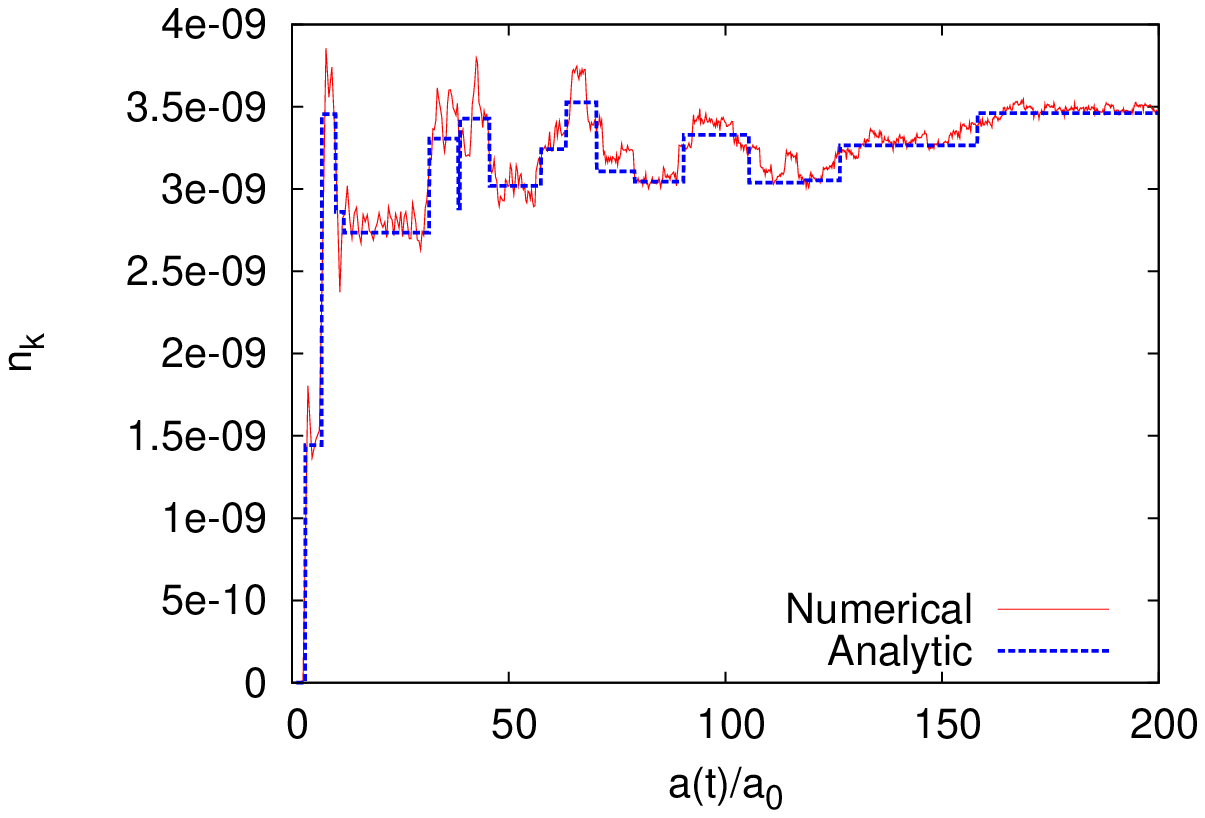,width=4in}{The time-dependence of the co-moving occupation number $n_k$ of $\chi$ particles with momentum $k$, produced via perturbative 4-legs interactions.
We have plotted both the approximate analytical estimate~\eqref{eqn:numk} (the dashed curve) and also the result of an exact numerical computation (the solid curve) of the co-moving mode functions $f_k(\tau)$.
We take $N=20$ inflaton fields with masses ranging from $0.01m$ to $10m$.  For simplicity, we assume equal initial amplitudes $\phi_{0,i} = \phi_0$ at the onset of reheating and also
equal couplings $g_i^2 = g^2$.  The parameters are $g^2 \phi_0^2 / m^2 =0.05$ and $k^2 / m^2 = 1000$.\label{fig:4legpertgraph}}

\subsection{Boltzmann Equation}

The energy density is
\begin{align}
a^4\rho_\chi &\approx \frac{1}{128\pi} \bigg\{\sum_i \int_{m_ia_0}^{m_ia(t)} d k
\frac{g_i^4 \phi_{0,i}^4}{a_{i,k}^3H_{i,k}} \\ \nonumber
&\qquad +2\sum_i \sum_{j>i}\int_{m_j a_0}^{m_ia(t)} d k \frac{g_i^2\phi_{0,i}^2g_j^2\phi_{0,j}^2}{\sqrt{a_{i,k}^{3}H_{i,k}a_{j,k}^{3}H_{j,k}}}
\cos(k[\psi_i^{+}(t_{i,k})-\psi_j^{+}(t_{j,k})]) 
\bigg\} \, .
\end{align}
The rate of change of the radiation energy density is
\begin{eqnarray}
&& a^{-4}\frac{\ud}{\ud t}(a^4\rho_{\chi})\cong \frac{1}{128\pi a^6}
\sum_i\bigg\{g_i^4\phi_{0,i}^4m_i \label{eqn:chienratefull} \\
&& + 2\sum_{j>i}g_i^2g_j^2\phi_{0,i}^2\phi_{0,j}^2m_i\left(\frac{m_j}{m_i}\right)^{3/2}\sqrt{\frac{H(a)}{H(m_ia/m_j)}}
\cos\left[k(\psi_i^+(\tau(a))-\psi_j^+(\tau(m_ia/m_j))\right] \, .
\bigg\} \nonumber
\end{eqnarray}
As in the single field case, the cross-terms in~\eqref{eqn:chienratefull} arise from QM interference between the decay channels.
However, unlike section \ref{sec:pert}, in this case the interference it is due to coherence between the \emph{same} decay channel 
(\emph{i.e.} pair annihilations) for \emph{different} inflaton fields.  On the other hand, in the single field case the interference
was between two different decay channels for the same inflaton.

Assuming a non-degenerate mass spectrum and averaging over a sufficiently long time interval for the interference term to vanish, we find
\begin{equation}
\label{boltz4legs}
  a^{-4} \frac{d}{dt} (a^4 \rho_\chi) \cong 2 \sum_i \frac{\left[ \sigma_{\phi_i\phi_i\rightarrow\chi\chi} v \right]_{v=0}}{m_i} \rho^2_{\phi_i} \, ,
\end{equation}
where we have used the fact that the energy density of the $i$-th inflaton is given by 
\smash{$\rho_{\phi_i}=\frac{1}{2}(\dot{\phi_i}^2 + m_i^2\phi_i^2) \cong \frac{m_i^2\phi_{i_0}^2}{2 a^3}$}.  
By comparison of (\ref{boltz4legs})
to (\ref{eqn:chienratefull}) we can extract the pair annihilation cross section for two $\phi_i$'s into a pair
of $\chi$ particles at zero relative velocity is
\begin{equation}
\left[\sigma_{\phi_i\phi_i \to \chi\chi}v\right]_{v=0}
= \frac{g_i^4}{64\pi m_i^2} \, .
\label{eqn:xsec}
\end{equation}
This is consistent with the perturbative QFT result one would obtain for pair annihilations of inflaton \emph{particles}, provided one
accounts for the subtlety concerning the limit $v\rightarrow 0$ discussed in section \ref{sec:pert}.

There is an interesting subtlety associated with the time-averaging that leads to (\ref{boltz4legs}).
For mass spectra which are very nearly degenerate (\emph{i.e.} $m_{\mathrm{max}}^2-m_{\mathrm{min}}^2 \ll m_{\mathrm{avg}}^2$)
the time over which we must average may be much longer than $m_i^{-1}$ and at any given time the oscillating terms in~\eqref{eqn:chienratefull} may contribute more
than the non-oscillating terms.  This is not surprising, since over time intervals much less than the inverse mass splittings
between the various inflatons, we may expect the many individual inflatons to behave as a single inflaton with oscillation
amplitude $\bar{\phi} \sim \sum_i \phi_i$.  In this case, our effective resonance parameter will be $\sum_i q_i$ where the sum is over those inflatons whose masses are
nearly degenerate, at least at early times before the inflatons have a chance to de-phase.
If almost all of the inflatons are degenerate in mass, it is possible for this effective resonance parameter to become large and the resonance to be efficient.
As a simple example, if all the inflatons have equal mass and the same initial conditions, then all of the cosines
will be $1$ and the ``oscillating'' terms (which there are $N^2-N$ of) 
will contribute more than the ``non-oscillating'' terms (which there are $N$ of).
Note that when we say degenerate, we mean that the mass splittings between \emph{all} pairs of inflatons are small,
not just the mass splittings between $m_i$ and $m_{i+1}$.

Assuming that the quantum mechanical interference terms are negligible, we can proceed to write down the effective Boltzmann equation for the inflaton oscillations.  Conservation of energy-momentum gives us
\begin{equation}\label{conservation}
a^{-3}\frac{d}{dt}\left(a^3\sum_i\rho_{\phi_i}\right)+ a^{-4}\frac{d}{dt}(a^4\rho_\chi)=0 \ .
\end{equation}
From (\ref{conservation}) and (\ref{boltz4legs}) we have the Boltzmann equation
\begin{equation}
\label{boltz4legs2}
   \frac{d}{dt} \left(\sum_i a^3 \rho_{\phi_i}\right) \cong -2 \sum_i \frac{\left[ \sigma_{\phi_i\phi_i\rightarrow\chi\chi} v \right]_{v=0}}{m_i a^3 } (a^3 \rho_{\phi_{i}})^2 \, .
\end{equation}
By direct integration one may verify that
\begin{equation}
  a^3 \sum_i \rho_{\phi_i} \rightarrow \mathrm{const}
\end{equation}
as $t \rightarrow \infty$.  Hence, the decay of the N-flatons does not complete: some finite co-moving energy density always freezes in.
This is quite analogous to the result for single field inflation with 4-legs interactions, see section \ref{sec:pert}. 
We conclude that the model (\ref{L2}) with $\sigma_i=0$ is not, in general, a viable scenario.

\section{Perturbative Inflaton Decays via Three-Leg Interactions}
\label{sec:3legs}

The 4-legs interactions discussed in the last section do \emph{not} permit a complete decay of the inflaton.  Therefore we must include other
couplings between $\phi_i$ and matter.  A natural candidate is the tri-linear (3-legs) interaction $\phi_i\chi^2$.  In this section we develop the theory
of perturbative reheating after inflation for the model (\ref{L2}) with $g_i^2=0$ but $\sigma_i \not= 0$.
For the nonperturbative decays, see section~\ref{sec:preheating}.

\subsection{Bogoliubov Calculation}

Let us now discuss the perturbative decays of the inflatons in the presence of 3-leg interactions $\sum_i \sigma_i^2 \phi_i \chi^2$.
Such perturbative decays are relevant during the final stages of reheating, or in the case where the inflaton oscillations at the end of
inflaton are small.  The quantum eigenmodes of the $\chi(t,{\bf x})$ field obey the oscillator-like equation (\ref{oscillator}) with effective frequency
\begin{eqnarray}
  \omega_k^2(\tau) &=& k^2 + a^2 \sum_i \sigma_i \phi_i(\tau) - \frac{a''}{a} \nonumber \\
                               &\cong& k^2 + \frac{1}{a^{1/2}(\tau)}\sum_i \sigma_i \phi_{0,i} \sin\left[m_i t(\tau) + \theta_i \right] \, , \label{omega3legs}
\end{eqnarray}
where on the last line we have used (\ref{Nosc}) and again neglected the $a'' / a \sim a^2 H^2$ term.

Following our previous formalism, we wish to evaluate the integrals in (\ref{beta_k}) for the relevant effective frequency (\ref{omega3legs}). 
The dominant contribution to the integral for the $i$-th inflaton comes 
from the moment when $2 k = a m_i$, corresponding to energy conservation for the process $\phi_i\rightarrow \chi\chi$.  We find the following result for the occupation number:
\begin{align}
\label{eqn:trilinpert}
n_{\bf k} = |\beta_k|^2= \frac{\pi}{16 k^3}&
\sum_i\bigg[\frac{\sigma_i^2\phi_{0,i}^2}{H_{i,k}}\Theta(a_{i,k}-a_0)\\
&+2\sum_{j>i}\frac{\sigma_i\phi_{0,i}\sigma_j\varphi_{0,j}}{\sqrt{H_{i,k}H_{j,k}}}
\cos(\psi_i(t_{i,k})-\psi_j(t_{j,k}))\Theta(a_{j,k}-a_0)\bigg]\Theta(a-a_{i,k}) \, , \nonumber
\end{align}
where $\psi_i(t) = -2k\tau(t)+m_it+\theta_i$, $m_ia_{i,k}=m_ia(t_{i,k})=2k$, $H_{i,k}=H(t_{i,k})$ and $m_i < m_j$ for $i<j$.

The accuracy of this analytic calculation is illustrated in Figure~\ref{fig:pertgraph}.
The oscillations appearing on the graph arise from boundary terms that have been ignored
in the stationary phase approximation (\emph{i.e.} higher order terms in the complete
asymptotic expansion of the phase integral~\eqref{beta_k}).
Physically, the variation of the background value of the inflaton field leads to 
continual creation and destruction of $\chi$ particles.
However, except in the vicinity of the stationary points $2k=m_ia$,
the destruction of particles tends to periodically balance the creation of particles,
leading to the oscillatory behaviour superimposed onto our analytic solution.

\EPSFIGURE[ht]{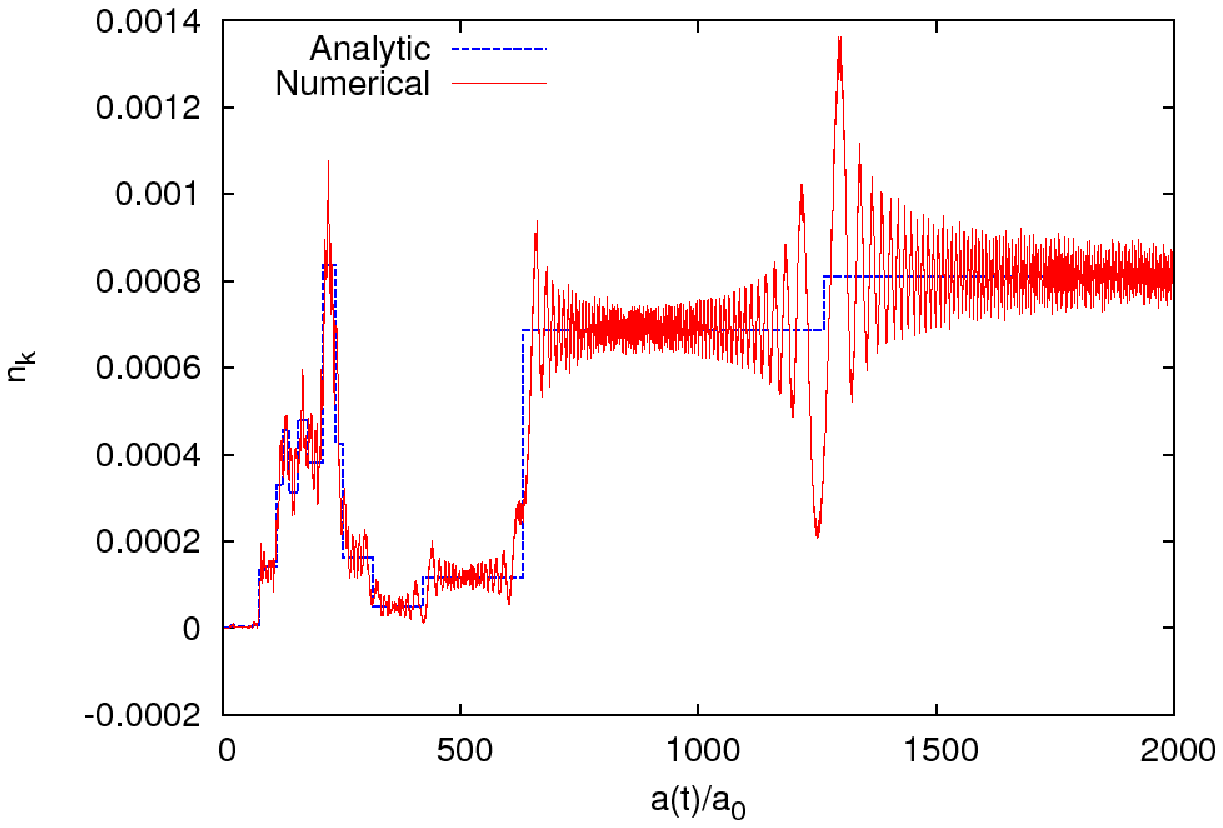,width=4in}{The time-dependence of the co-moving occupation number $n_k$ of $\chi$ particles with momentum $k$, produced via perturbative tri-linear interactions.
We have plotted both the approximate analytical estimate~\eqref{eqn:trilinpert} (the dashed curve) and also the result of an exact numerical computation (the solid curve) of the co-moving mode functions $f_k(\tau)$.
We take $N=20$ inflaton fields with masses ranging from $0.01m$ to $10m$.  For simplicity, we assume equal initial amplitudes $\phi_{0,i} = \phi_0$ at the onset of reheating and also
equal couplings $\sigma_i = \sigma$.  The parameters are $\sigma \phi_0 / m^2 =0.05$ and $k^2 / m^2 = 1000$.  Compared with the 4-legs interaction in Fig.~\ref{fig:4legpertgraph}, 
the particle occupation numbers are greater by a factor of $a_{ik}^3 \sim (k/m)^{3/2}  \sim 10^5$ as expected from comparison of~\eqref{eqn:numk} and~\eqref{eqn:trilinpert}.\label{fig:pertgraph}}

\subsection{Boltzmann Equation}

We find the average rate of radiation  energy production
\begin{equation}\label{width}
a^{-4}\frac{d}{dt}({a^4\rho_{\chi}})=
\frac{1}{a^3}\sum_i\frac{\sigma_i^2\phi_{0,i}^2m_i}{64\pi}
=\frac{1}{a^3}\sum_i \Gamma_{\phi_i\rightarrow\chi\chi} \rho_{\phi_i} \ , 
\end{equation}
where
\begin{equation}
\Gamma_{\phi_i\rightarrow\chi\chi} = \frac{\sigma_i^2}{32\pi m_i} \ .
\end{equation}
Here we recognize $\Gamma_{\phi_i\rightarrow\chi\chi}$ as the decay width for $\phi_i \to \chi\chi$ due to the coupling 
\smash{$\frac{1}{2}\sigma_i\phi_i\chi^2$}.  This coincides with the result one would compute in perturbation theory for the decay
of inflaton particles.

Conservation of energy gives $a^{-4}\partial_t(a^4 \rho_\chi) = -a^{-3}\sum_i\partial_t(a^3 \rho_{\phi_i})$, hence we can write (\ref{width}) as a series of $N$ decoupled
equations
\begin{equation}
  \frac{d}{dt}(a^3\rho_{\phi_i}) = -\Gamma_{\phi_i\rightarrow\chi\chi} (a^3 \rho_{\phi_i}) \, .
\end{equation}
By direct integration we see that the co-moving energy density of each inflaton drops exponentially $a^3\rho_{\phi_i} \sim e^{-\Gamma_i t}$ and hence the decay process does complete.  This is similar to the single field
case with 3-legs interactions, but unlike the case with 4-legs interaction (both single and multi-field versions).

\section{Reheating into the Standard Model}
\label{sec:SM}

\subsection{Coupling to Gauge Fields}

The prototype Lagrangian (\ref{L2}) allows us to study the decay of the inflaton condensate into scalar field fluctuations from a phenomenological perspective.  
However, in order to truly make contact with the usual picture of the hot big bang, we would like to understand in detail the production of standard model degrees of freedom at the end of inflation.  
A novel feature of string theory inflation model-building is that it is possible, at least in principle, to determine such couplings from a bottom-up perspective.
As a first step to understanding reheating of the SM after N-flation, we consider a $U(1)$ gauge field $A_\mu$ with field strength $F_{\mu\nu} = \partial_\mu A_\nu - \partial_\nu A_\mu$.
Since $\phi_i$ are axions, we expect couplings to $F\wedge F$.  Hence, we consider the action
\begin{equation}
\label{axion_int}
  \mathcal{L} =  \sum_{i=1}^N \left[ -\frac{1}{2}\partial_\mu \phi_i \partial^\mu \phi_i - \frac{m_i^2}{2}\phi_i^2  \right]
    -\frac{1}{4}F_{\mu\nu}F^{\mu\nu} - \sum_i\frac{\phi_i}{ 4 M_i} F^{\mu\nu}\tilde{F}_{\mu\nu} \, ,
\end{equation}
where $\tilde{F}^{\mu\nu} = \frac{1}{2}\epsilon^{\mu\nu\rho\sigma}F_{\rho\sigma}$ is the dual field strength and
$M_i$ is some symmetry breaking scale.  Typically $M_i \sim f_i$, however, we do not require this for our analysis.

It is worth considering how (\ref{axion_int}) arises in the IIB string theory model discussed above.  Since the axions $\tilde{\phi}_i$ are associated with 4-cycles, it is natural to consider localizing the SM on a D7-brane
wrapping the $j$-th cycle.  (See also \cite{BBHK,green}.)  In this case one expects a coupling of the form 
\begin{equation}
\mathcal{L}_{\mathrm{int}} = - \lambda \tilde{\phi}_j F \tilde{F}
\end{equation}
to brane-bound gauge fields.  In general, the mass basis $\phi_i$ will not be aligned with the basis $\tilde{\phi}_i$ associated with the geometrical 4-cycles.  Hence, a rotation in field space 
(\ref{field_rot}) will be necessary to put the axion action (\ref{non_can}) in the canonical form (\ref{axion_int}).  For a general rotation \emph{all} the mass basis axions $\phi_i$ will appear 
in the sum (\ref{field_rot}) and, accordingly, all of the N-flatons are expected to couple to $F \wedge F$.  This argument leads directly to an action of the form (\ref{axion_int}) where the couplings $M_i$
are related to the coefficients $a^{(i)}_{j}$ in the field-space rotation (\ref{field_rot}).  These are, obviously, very model dependent.  Normalizing $\sum_j \left( a_{j}^{(i)} \right)^2 = 1$ we expect $a_{j}^{(i)} \sim N^{-1/2}$ 
so that $M_i \sim N^{1/2} / \lambda$.

Note that the argument above works equally well for gauge fields living on \emph{any} D7-brane wrapping \emph{any} 4-cycle of the 
Calabi-Yau compactification.  In general, there might be many such branes so that the compactification contains the usual SM
in addition to a large number of hidden sectors.  Those axions which are still dynamical at the end of inflation will reheat into \emph{all} sectors more-or-less equally 
(of course, in a specific model this depends on the details of the field rotation and the placement of branes in the Calabi-Yau manifold).  Relics in the hidden sectors might be dark matter candidates,
or they might pose cosmological problems, depending on model parameters.  See \cite{green} for more details.

The equation of motion for the gauge field derived from (\ref{axion_int}) is
\begin{equation}
  \nabla_{\mu} F^{\mu\nu} + \sum_i \frac{1}{M_i} (\nabla_\mu\phi_i) \tilde{F}^{\mu\nu} = 0 \, .
  \label{eqn:modefxn}
\end{equation}
As usual, we study the linear quantum fluctuations of the matter field, treating the inflaton oscillations (\ref{Nosc}) as a classical background.
We work in a transverse gauge, with $A^0=0$ and $\del_i A^i = 0$, and decompose into circularly polarized mode functions
\begin{equation}
  A^i(t,{\bf x}) = \int \frac{d^3k}{(2\pi)^{3/2}} \sum_{\lambda=\pm} \left[ e_\lambda^i({\bf k}) a_{{\bf k}}^\lambda e^{i {\bf k}\cdot {\bf x}} A^\lambda_{k}(t) + \mathrm{h.c.} \right] \, ,
\end{equation}
where the circular polarization vectors satisfy ${\bf k} \times {\bf e}_{\pm} = \mp \, i \, k \, {\bf e}_{pm}$,
$a_{\bf k}^\lambda$ are q-number annihilation/creation operators and ``h.c.'' denotes the Hermitian conjugate of the preceding term.
Using this expansion,~\eqref{eqn:modefxn} gives
\begin{equation}
\label{eqn:Amode}
  \ddot{A}_k^{\pm} + H \dot{A}_k^{\pm} +\left[ \frac{k^2}{a^2} \mp \frac{k}{a}\sum_i \frac{\dot{\phi}_i(t)}{M_i}  \right] A_k^{\pm} = 0 \, .
\end{equation}

\subsection{Nonperturbative Preheating}

To investigate the possibility of nonperturbative photon production, we would like to put (\ref{eqn:Amode}) in the form of the QP Mathieu equation.
To this end we introduce the co-moving modes $\tilde{A}^{\pm} = a^{1/2}A^{\pm}$ and define a new time variable $2z = mt$ where $m$ is a typical inflaton mass.
Treating the expansion of the universe adiabatically one
arrives at the QP Mathieu form (\ref{hill_3}) with parameters
\begin{equation}
A_k = \frac{4k^2}{m^2a^2},
\qquad
q_{i}^{\pm} = \pm\frac{2k m_i \phi_{0,i}}{m^2 M_i a^{5/2}} \, .
\end{equation}
We can now employ our results from subsection \ref{subsec:QP}.
Notice that, unlike the scalar field cases studied in section \ref{sec:preheating}, both our $A_k$ and $q_{i}$ parameters now depend on $k$.
The resonance parameters for the mode functions are no longer initially distributed along a vertical line,
but instead along the parabola
\begin{equation}
\label{eqn:Astart}
A_k^0 = \left(\frac{m}{m_i}\frac{M_i}{\phi_{0,i}}\right)^2 \left(q_{i}^0\right)^2 
= \left(\sum_i \frac{m_i\phi_{0,i}}{m M_i}\right)^{-2}\left(\sum_i q_i^0\right)^2 \ .
\end{equation}
The expansion of the universe causes these parameters to ``flow'' along the lines
\begin{equation}
\label{eqn:Acurves}
A_k =  A_k^0 \left(\frac{q_{i}}{q_{i}^0}\right)^{4/5} = A_k^0\left(\frac{\sum_i q_i}{\sum_i q_i^0}\right)^{4/5}\, .
\end{equation}
Hence, all of the field modes will trace out curves in the $(A,{\bf q})$ plane that lie above 
\smash{$A_k = \left(\sum_i\frac{m_i}{m}\frac{\phi_{0,i}}{M_i}\right)^{-2}(\sum_i q_{i})^2
\equiv C^{-2}(\sum_i q_{i})^2$}.
This is illustrated in Fig.~\ref{fig:stabgauge} for several choices of the prefactor $C$.
In Fig.~\ref{fig:floqgauge} we show the Floquet exponents along several of these initial curves, 
which provides a measure of the initial instability of the gauge fields.

\EPSFIGURE[ht]{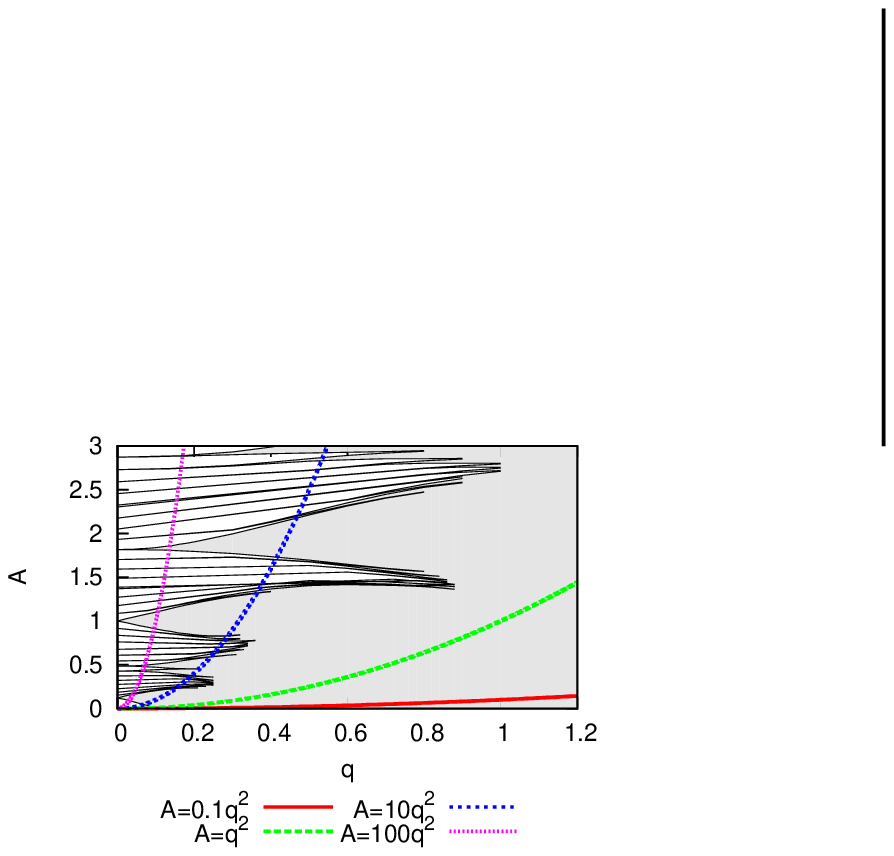,width=4in}{The permitted regions for gauge field modes in the $(A,q)$ plane.
The field modes must lie above the curves $A=C^{-2}(2q)^2$, where the 
parameter $C = \sum_i m_i\phi_{0,i} / mM_i$.\label{fig:stabgauge}}

\EPSFIGURE[ht]{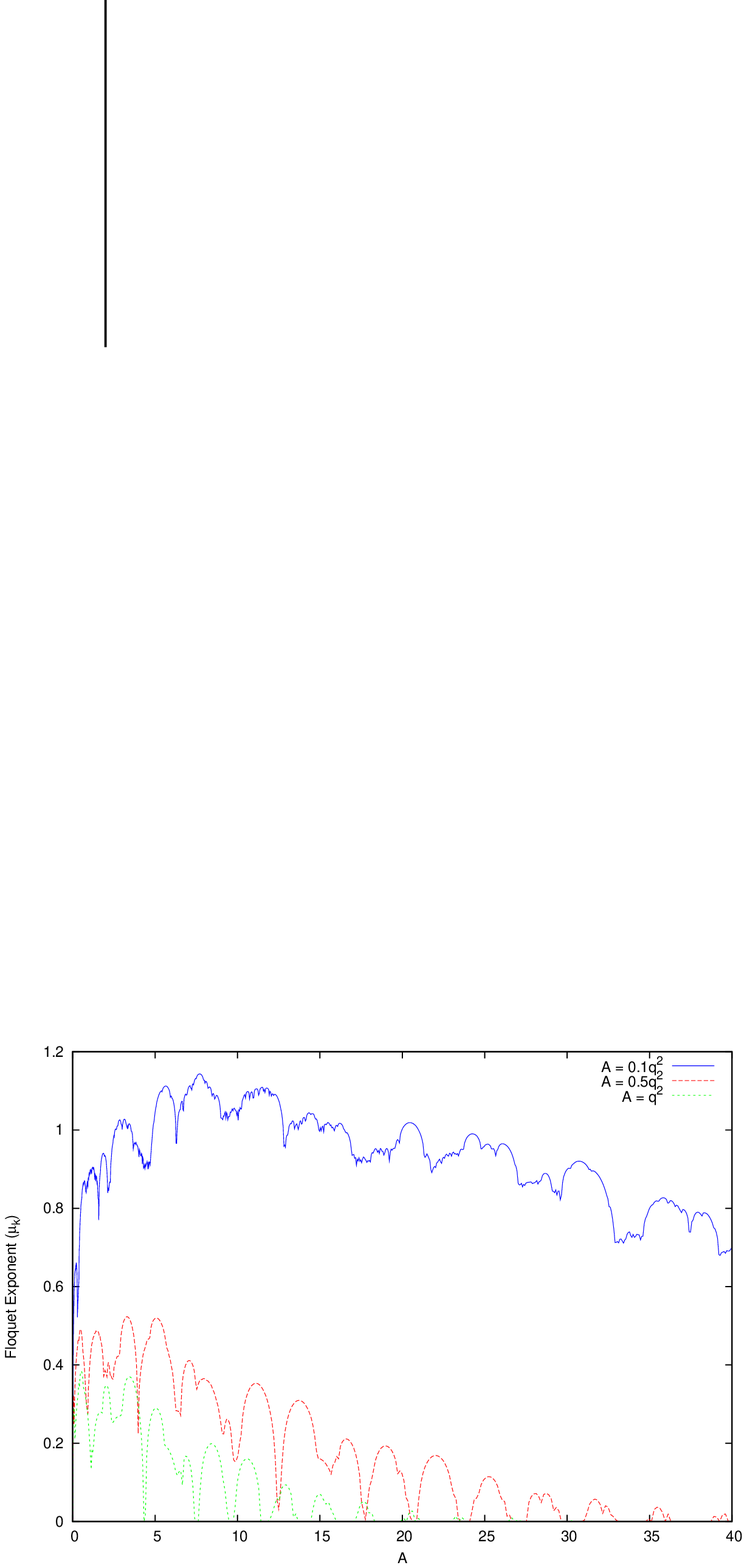,width=4in}{Floquet exponents along the curves $A=C^{-2}(2q)^2$ for several choices of $C$ and the
same parameters as in Figure~\ref{fig:stabgauge}.\label{fig:floqgauge}}

For $q_i$ values such that the individual bands are distinct, the condition $q_i^{\mathrm{eff}} \gg 1$
while the mode is in the $i$th first order resonance band at $A = (m_i/m)^2$ translates into
\begin{equation}
\label{eqn:gaugerescond}
\left(\frac{M_i}{\phi_{0,i}}\right)^2 \ll 1 \ .
\end{equation}
This is the straightforward generalization of the condition found in~\cite{ref:preheatderiv}.
Meanwhile, in order to for the parabola to pass through the region where many of the first order bands
have overlapped and we again expect the resonance to be strong, we instead require
\begin{equation}
   \frac{C m_{\mathrm{max}}^{\mathrm{osc}} m}{N \delta \bar{m^2}} = \sum_i \frac{m_{\mathrm{max}}^{\mathrm{osc}}m_i \phi_{0,i}}{N_{\mathrm{osc}}\delta \bar{m^2}M_i} \gg 1
  \label{eqn:gaugerescond2}
\end{equation}
where we have again assumed that all of the squared mass splittings $\delta \bar{m^2}$ and $q_i=\bar{q}$'s are equal.

It is worth considering whether the condition for an efficient resonance is satisfied in the IIB string theory construction discussed in section \ref{sec:Nflation}.
As mentioned above, we expect a coupling $\tilde{\phi}_j F \tilde{F} / M$ to brane-bound fields living on the $j$-th cycle.  In string theory, we typically have $M$ below
the Planck scale, but not significantly so.  Thus, we assume $M \sim M_p$.  Upon rotating to the interaction basis we have a coupling to all of the N-flatons $\phi_i$.  
For a generic rotation (\ref{field_rot}) we have $a_j^{(i)} \sim N^{-1/2}$ (as discussed above) which leads to the estimate $M_i \sim \sqrt{N} M_p$ for the couplings in (\ref{axion_int}).  
Combining this with the estimate $\phi_{0,i} \sim M_P/\sqrt{N_{\mathrm{osc}}}$ 
gives $\phi_{i,0}/M_i \sim 1/\sqrt{NN_{\mathrm{osc}}} \sim \sqrt{10}/N$.  In this case the condition~\eqref{eqn:gaugerescond} for a single one of the fields to provide an efficient resonance is \emph{not} satisfied.
Meanwhile, in order for us to be in the regime~\eqref{eqn:gaugerescond2} where many resonances have overlapped we require
\begin{equation}
\sum_i \frac{m_{\mathrm{max}}^{\mathrm{osc}}m_i \phi_{0,i}}{N\delta\bar{m^2}M_i} \sim \frac{N_{\mathrm{osc}} m_{\mathrm{typical}}m_{\mathrm{max}}^{\mathrm{osc}}}{N\delta\bar{m^2}}\frac{\sqrt{10}}{N} \sim \frac{m_{\mathrm{typical}}^2}{N\delta\bar{m^2}} \gg 1 \, .
\end{equation}
Taken together, these constraints imply that, as long as the mass spectrum is not too degerate, nonperturbative effects will be weak and the decay of the inflaton will be dominated by perturbative processes
$\phi_i \rightarrow\gamma\gamma$.  Obviously, this conclusion will depend considerably on the model-building detail, especially the field rotation (\ref{field_rot}) and the presence of hidden sectors in the Calabi-Yau compactification volume.

\subsection{Perturbative Reheating}

With the expectation for perturbative decays described above, we now investigate this case.
Transforming~\eqref{eqn:Amode} to conformal time gives
\begin{equation}
\frac{\ud^2}{\ud \tau^2} {A}_{k}^{\pm} 
+ \left(k^2 \mp ka\sum_i\frac{\dot\phi_i}{M_i}\right)A_{k}^\pm
    = 0 \, .
\end{equation}
We see that our modes functions $A_{k}^{\pm}$ are already co-moving
with an effective frequency given by
\begin{equation}
(\omega_k^\pm)^2 = k^2 \mp 
\frac{k}{\sqrt{a}} \sum_i\frac{\phi_{0,i}m_i}{M_i}
\left[\cos(m_it+\theta_i)
- \frac{3H}{2m_i}\sin(m_it + \theta_i)\right] \, .
\end{equation}
The second term in the square brackets is small compared to the first term 
during reheating and can be ignored.
Following our procedure from before (replacing $f_{k}$ with $A_{k}^\pm$, 
and assuming $\sum_i \frac{m_i \phi_{0,i}}{m_{\mathrm{max}}M_i} \ll 1$),
we find for the co-moving number density (ignoring the cross terms)
\begin{equation}
n_{k}^\pm =  
\frac{\pi}{64 k^3} \sum_i \frac{m_i^4\phi_{0,i}^2}{H_{i,k}M_i^2}
=\frac{\pi\sqrt{12}}{32} \frac{1}{k^{3/2}}
\left(\sum_jm_j^2\phi_{0,j}^2\right)^{-1/2}
\sum_i \frac{\phi_{0,i}^2m_i^{5/2}}{M_i^2} \, ,
\end{equation}
where we have assumed that the energy density of the universe is
dominated by the inflatons (\ie $H^2\cong\frac{1}{6a^3M_p^2}\sum_i{m_i^2\phi_{0,i}^2}$)
in the last step.
Accounting for both polarizations and time averaging, the gauge field gains energy at a rate 
\begin{equation}
\label{gauge_boltz}
a^{-4}\frac{d}{d t}(a^4 \rho_{A})
= \frac{2}{128\pi a^3} \sum_i \frac{m_i^5 \phi_{0,i}^2}{2 M_i^2}
= \sum_i \Gamma_{\phi_i\rightarrow\gamma\gamma} \rho_{\phi_i} \, .
\end{equation}
The decay rate for $\phi_i$ is given by
\begin{equation}
\Gamma_{\phi_i\rightarrow\gamma\gamma} = \frac{m_i^3 }{64 \pi M^2_i} \, .
\end{equation}
This agrees with the more familiar result for the perturbative decay of inflaton \emph{particles}.

As before, we can use conservation of energy to re-write (\ref{gauge_boltz}) as
\begin{equation}
  \frac{d}{dt}(a^3 \rho_{\phi_i}) = - \Gamma_{\phi_i\rightarrow\gamma\gamma} (a^3 \rho_{\phi_i})
\end{equation}
so that $a^3 \rho_{\phi_i} \sim e^{-\Gamma_i t} \rightarrow 0$ as $t \rightarrow \infty$, indicating that
the N-flatons are permitted to decay completely into photons.

Let us now estimate the reheat temperature provided by the above decays.
When decaying into a single field, we have the standard estimate for the reheat temperature $T_R$~\cite{KLS97}
\begin{equation}
  T_R \sim \left(\frac{90}{g\pi^2}\right)^{1/4}\sqrt{M_P\Gamma} \, 
\end{equation}
where
\begin{equation}
  \Gamma = \sum_i \Gamma_i \frac{\rho_{\phi_i}}{\rho_{\phi}}
\end{equation}
and $g$ is the number of relativistic degrees of freedom (including $A$) which interact
with our ``photons''.
Let's assume that the inflatons all have similar masses and oscillation amplitudes.
Then $\rho_{\phi_i}/\rho_{\phi} \sim N_{\mathrm{osc}}^{-1}$ and we have
\begin{equation}
  T_R \sim \left(\frac{90}{g\pi^2}\right)^{1/4}\sqrt{\frac{M_P}{64\pi}\sum_i{\frac{m_i^3}{N_{\mathrm{osc}}M_i^2}}}
    \sim \sqrt{\frac{m_i^3}{NM_p}} \sim \frac{10^{10}\mathrm{GeV}}{\sqrt{N}} \, ,
\end{equation}
where we used $m_i \sim 10^{-6}M_P$ for Nflation.
Notice that the suppression factor is the total number of inflatons
(which determines the rotation in the field space), not the number of inflatons contributing
to the reheating.

\section{Applications to More General Models}
\label{sec:generalizations}


\subsection{Generalizations}

Clearly, much of our analysis in this paper is not specific to N-flation \cite{Nflation} or its IIB string theory realization \cite{matrix_spectrum}.  The general form of the 
potential (\ref{V_i}) will hold near the minimum in a large class of models.  Thus, the decay of the coherent oscillations (\ref{Nosc})
might be relevant for a variety of multi-field constructions.  In this subsection we speculate on some possible applications of our results
outside of the context of N-flation.\footnote{Note, however, that the following discussion will 
certainly not apply in \emph{all} models.  For 
example, in the presence of steep tachyonic directions the energy of the inflaton will be mostly drained by spinodal decomposition 
\emph{before} the oscillations 
(\ref{Nosc}) can set in \cite{tac}.  Also, it may happen that the potential has a very steep minimum so that the form (\ref{V_i}) is not valid for the entire
duration of a single oscillation, as in the roulette inflation model \cite{BBHK}.}  We leave a detailed discussion to future works.

In particular, the decay of the inflaton oscillations (\ref{Nosc}) via couplings such as (\ref{L2}) and (\ref{axion_int}) might be important for
models of multi-field inflaton on the string theory landscape \cite{multi_land,staggered,rand1,rand2,stream}.  The inflationary trajectory 
across the cosmic landscape typically involves a large number of twists, turns and bifurcations, leading to various features \cite{rand1,rand2}
and nongaussianities \cite{stream} in the Cosmic Microwave Background (CMB) fluctuations.   Inflation might end when this trajectory encounters
a steep minimum, leading to a post-inflationary evolution where are large number of scalars $\phi_i$ oscillate according to (\ref{Nosc}).  Depending
on the specific corner of the landscape, the mass spectrum $m_i$ and initial conditions at the onset of reheating might be
more-or-less randomly distributed.  

Finally, our analysis might also have some relevance for the model of \cite{vikman} if the final stage of inflation is
not entirely along the inflaton direction associated with our visible standard model.  In that case our results could have relevance
for the nonperturbative production of hidden sector baryonic dark matter.  See also \cite{yuki} for a related discussion.

\subsection{Modular Inflaton Couplings to Gauge Fields}

A more specific stringy scenario where our analysis might prove relevant is racetrack inflaton \cite{race1,race2}.  In the ``improved'' construction
of \cite{race2} inflation involves the complex motion of several moduli and axions in the IIB KKLT vacua discussed above.  If there is a prolonged
phase of oscillations about the minimum\footnote{Such a phase \emph{does} exist in the case of the original racetrack model \cite{race1}
where nonperturbative preheating effects are absent in the scalar sector \cite{zhiqi}.}, then our results will apply.  
We expect couplings of the form (\ref{axion_int}) for the inflatons associated
with axions arising from wrapping the 4-form field on various 4-cycles of the Calabi-Yau compactification manifold.  However, there will also
be couplings between brane-bound gauge fields and the moduli which control the volume of these 4-cycles.  Such couplings typically take the form
$\tilde{\phi}_j F^2$ for gauge fields living on the $j$-th 4-cycle \cite{BBHK}.  On rotating to the mass basis (\ref{field_rot})
we will have couplings between many inflatons and the field strength squared.  Thus, in addition to the axion coupling (\ref{axion_int}), there is a 
phenomenological motivation to consider ``modular'' couplings of the form
\begin{equation}
  \mathcal{L} =  \sum_{i=1}^N \left[ -\frac{1}{2}\partial_\mu \phi_i \partial^\mu \phi_i - \frac{m_i^2}{2}\phi_i^2  \right]
  -\frac{1}{4}F_{\mu\nu}F^{\mu\nu} - \sum_i\frac{\phi_i}{ 4 M_i} F^{\mu\nu} F_{\mu\nu} \, .
  \label{eqn:lang_moduli}
\end{equation}
Here we assume $\sum_i \phi_i(t)/M_i \ll 1$ in order to justify the neglect of higher order terms in the 
inflaton potential and of dimension-6 and higher interaction terms in (\ref{eqn:lang_moduli}).
Interactions of the type considered in (\ref{eqn:lang_moduli}) might play a role in a model of assisted inflation from string theory using the dynamics of 
moduli and, perhaps, also in the model of \cite{berglund}.  Working in transverse gauge, we find that our mode functions obey
\begin{equation}
  \ddot{A}_{\bf k}^{\pm} + \left[ H + \sum_i \frac{\dot{\phi}_i(t)}{M_i} \right] \dot{A}_{\bf k}^{\pm} + \frac{k^2}{a^2}A_{\bf k}^{\pm} = 0 \, .
\end{equation}
Here we work to leading order in $\sum_i \phi_i(t)/M_i \ll 1$.  To put this into the form of a time-dependent frequency oscillator, we define a new field 
$\tilde{A} = \sqrt{a \tau} A$.  The equation for our mode function now becomes
\begin{equation}
  \label{eqn:moduli_mode}
  \ddot{\tilde{A}}_{\bf k}^{\pm} + \left[ \frac{k^2}{a^2} - \frac{1}{2}\sum_i \frac{\ddot{\phi}_i(t)}{M_i} - \frac{1}{2} H \sum_i \frac{\dot{\phi}_i(t)}{M_i} - \frac{1}{4} H^2 - \frac{1}{2} \dot{H}\right] \tilde{A}_{\bf k}^{\pm} = 0 \, .
\end{equation}
Again, we drop terms which are higher order in $\sum_i \phi_i(t) / M_i \ll 1$.  
The fact that the modular coupling in (\ref{eqn:lang_moduli}) must be treated as a small perturbation 
to the gauge field kinetic term rules out strong nonperturbative preheating effects during the stages when the $\sum_i m_i^2\phi_i^2$ form of the potential is valid.  Hence, we study instead the perturbative decays of the coherent inflaton oscillations into gauge field quanta.
Performing our standard procedure as in section \ref{sec:pert} and accounting for both photon polarizations we find the effective Boltzmann equation
\begin{equation}
  \frac{\ud}{\ud t}(a^3\rho_{\phi_i}) = -\Gamma^{\mathrm{mod}}_{i}(a^3\rho_{\phi_i})
\end{equation}
with decay rate
\begin{equation}
  \Gamma^{\mathrm{mod}}_{i} = \frac{m_i^3}{64\pi M_i^2} \, .
\end{equation}
This again matches the standard result for decays of inflaton \emph{particles}.  The interaction (\ref{eqn:lang_moduli}) permits the complete decay of the inflaton with a reheat temperature 
$T_R \sim N^{-1/2}10^{10}\mathrm{GeV}$.

\section{Conclusions}
\label{sec:conclusions}


In this paper we have studied (p)reheating after inflation in models with a large number of inflaton fields.  
Such models are quite natural from the string theory perspective and may predict significant gravitational waves
in a regime where the individual inflaton displacements are sub-Planckian, via the assisted inflation mechanism.
We have focused on N-flation \cite{Nflation}
in particular, however, we expect that our results may apply more generally to a variety of multi-field inflation models.  In particular, our finding may
have some relevance for multi-field inflation on the cosmic landscape \cite{rand1,rand2,stream}.  

We have proceeded phenomenologically, considering
a variety of possible couplings between the inflaton fields and matter.  We studied the decay of the N-flatons into scalar field matter via both 4-leg and 3-leg interactions.
In the former case, we found that reheating does not complete (similarly to single field inflation).  In the latter case, we found that reheating may complete
in the perturbative regime, and also that strong nonperturbative tachyonic resonance is possible for certain parameters.  

In the case of 3-leg interactions, the fluctuations of the preheat field display exponential growth in certain regions of phase space.  The 
structure of the stability/instability bands is quite rich and is related to the mathematical properties of  the quasi-periodic Mathieu equation.
We have shown that the presence of multiple frequencies in the effective mass of the preheat field leads to interference effects and the dissolution
of the stable regions.  

We also studied the decay of the N-flatons to gauge fields.  In the original proposal \cite{Nflation} the N-flatons are axions, therefore we considered
a coupling of the form $\sum_i \phi_i F \tilde{F}$.  We couple \emph{all} of the N-flatons to a single $U(1)$ gauge field since the mass basis $\phi_i$
and interaction basis $\tilde{\phi}_i$ are not, in general, aligned with one-another.  We have discussed how this interaction arises in the context of
type IIB string theory vacua.  We found that (p)reheating in this case proceeds very similarly to the 3-legs scalar interaction.  Reheating can complete
perturbatively and strong nonperturbative preheating effects are possible.

For completeness, we have also considered a coupling $\sum_i\phi_i F^2$ to gauge fields.  Such terms may arise in string theory models where assisted
inflation is realized from the dynamics of moduli fields.  

During the course of our analysis we have developed an analytical theory of the perturbative decays of coherent, classical inflaton oscillations.  This
formalism applies much more generally than the N-flation models considered here.  We have used this method to compare the difference between
the decay/annihilation of a classical, homogeneous inflaton condensate with the analogous process for inflaton \emph{particles}.  We have seen that
these two processes are essentially different.  In the case of a classical inflaton, the various decay channels of $\phi$ contribute coherently
to the amplitude.  Upon squaring this, we see that these decay channels can interfere quantum mechanically leading to oscillatory contributions
to the effective Boltzmann equation.  These oscillations disappear upon averaging over a sufficiently long time interval and we recover the usual results
for decay/annihilation of inflaton particles at rest.

There are a number of directions for future studies.  It would be interesting to consider a rigorous, stabilized stringy embedding of N-flation and identify
the location of the standard model and the various decay channels of the inflaton \cite{green}.  Using the results of this paper, one could assess the viability of such
a construction and study the production of exotic relics, gravitational waves \cite{gw}, nongaussianities \cite{ng1,ng2}, etc.  
It would also be interesting to study post-inflationary dynamics in racetrack inflation \cite{race1,race2}, or in a more general string landscape 
setting \cite{rand1,rand2}.

There does not yet exist a complete theory of preheating after multi-field inflation.  The result of this paper should be considered as a step
in the direction of a more comprehensive study.  As in the single field case, we expect that a rich spectrum of perturbative and nonperturbative
QFT effects are possible \cite{KLS97}, in addition to distinctively stringy processes \cite{BBC,KY,BBHK}.

\section*{Acknowledgments}
This paper is dedicated to the memory of Lev Kofman, who passed away before this project could be seen to completion.
J.B.~is supported by an NSERC Postgraduate Scholarship
and the University of Toronto. L.K.~was supported by NSERC and  CIFAR.
N.B.~is supported by CITA.
We thank Diana Battefeld, Thorsten Battefeld, Dick Bond, Jim Cline, Joseph Conlon, Gary Felder, Andrew Frey, John Giblin Jr.,  Zhiqi Huang, Renata Kallosh,
Andrei Linde, Liam McAllister, Sergei Prokushkin and Pascal Vaudrevange for useful discussions and correspondence.


\begin{thebibliography}{99}

\bibitem{kofman25plus}

  L.~Kofman,
  ``Preheating After Inflation,''
  Lect.\ Notes Phys.\  {\bf 738}, 55 (2008).

\bibitem{KLS94}

  L.~Kofman, A.~D.~Linde and A.~A.~Starobinsky,
  ``Reheating after inflation,''
  Phys.\ Rev.\ Lett.\  {\bf 73}, 3195 (1994)
  [arXiv:hep-th/9405187].

\bibitem{KLS97}

  L.~Kofman, A.~D.~Linde and A.~A.~Starobinsky,
  ``Towards the theory of reheating after inflation,''
  Phys.\ Rev.\  D {\bf 56}, 3258 (1997)
  [arXiv:hep-ph/9704452].

\bibitem{frolov}

  A.~V.~Frolov,
  ``Non-linear Dynamics and Primordial Curvature Perturbations from
  Preheating,''
  arXiv:1004.3559 [gr-qc].

\bibitem{ferm} 
  
  P.~B.~Greene and L.~Kofman,
  ``Preheating of fermions,''
  Phys.\ Lett.\  B {\bf 448}, 6 (1999)
  [arXiv:hep-ph/9807339].

  P.~B.~Greene and L.~Kofman,
  ``On the theory of fermionic preheating,''
  Phys.\ Rev.\  D {\bf 62}, 123516 (2000)
  [arXiv:hep-ph/0003018].

  M.~Peloso and L.~Sorbo,
  ``Preheating of massive fermions after inflation: Analytical results,''
  JHEP {\bf 0005}, 016 (2000)
  [arXiv:hep-ph/0003045].

\bibitem{res}

  P.~B.~Greene, L.~Kofman, A.~D.~Linde and A.~A.~Starobinsky,
  ``Structure of resonance in preheating after inflation,''
  Phys.\ Rev.\  D {\bf 56}, 6175 (1997)
  [arXiv:hep-ph/9705347].

\bibitem{tac}

  G.~N.~Felder, J.~Garcia-Bellido, P.~B.~Greene, L.~Kofman, A.~D.~Linde and I.~Tkachev,
  ``Dynamics of symmetry breaking and tachyonic preheating,''
  Phys.\ Rev.\ Lett.\  {\bf 87}, 011601 (2001)
  [arXiv:hep-ph/0012142].

  G.~N.~Felder, L.~Kofman and A.~D.~Linde,
  ``Tachyonic instability and dynamics of spontaneous symmetry breaking,''
  Phys.\ Rev.\  D {\bf 64}, 123517 (2001)
  [arXiv:hep-th/0106179].

\bibitem{tac_res}

  J.~F.~Dufaux, G.~N.~Felder, L.~Kofman, M.~Peloso and D.~Podolsky,
  ``Preheating with Trilinear Interactions: Tachyonic Resonance,''
  JCAP {\bf 0607}, 006 (2006)
  [arXiv:hep-ph/0602144].

\bibitem{BBC}

  N.~Barnaby, C.~P.~Burgess and J.~M.~Cline,
  ``Warped reheating in brane-antibrane inflation,''
  JCAP {\bf 0504}, 007 (2005)
  [arXiv:hep-th/0412040].

\bibitem{KY}

  L.~Kofman and P.~Yi,
  ``Reheating the universe after string theory inflation,''
  Phys.\ Rev.\  D {\bf 72}, 106001 (2005)
  [arXiv:hep-th/0507257].

\bibitem{BBHK}

  N.~Barnaby, J.~R.~Bond, Z.~Huang and L.~Kofman,
  ``Preheating After Modular Inflation,''
  JCAP {\bf 0912}, 021 (2009)
  [arXiv:0909.0503 [hep-th]].

\bibitem{stringy_reheating}

  N.~Barnaby and J.~M.~Cline,
  ``Creating the universe from brane-antibrane annihilation,''
  Phys.\ Rev.\  D {\bf 70}, 023506 (2004)
  [arXiv:hep-th/0403223].

  N.~Barnaby and J.~M.~Cline,
  ``Tachyon defect formation and reheating in brane-antibrane inflation,''
  Int.\ J.\ Mod.\ Phys.\  A {\bf 19}, 5455 (2004)
  [arXiv:hep-th/0410030].

  A.~R.~Frey, A.~Mazumdar and R.~C.~Myers,
  ``Stringy effects during inflation and reheating,''
  Phys.\ Rev.\  D {\bf 73}, 026003 (2006)
  [arXiv:hep-th/0508139].

  X.~Chen and S.~H.~Tye,
  ``Heating in brane inflation and hidden dark matter,''
  JCAP {\bf 0606}, 011 (2006)
  [arXiv:hep-th/0602136].

  A.~Berndsen, J.~M.~Cline and H.~Stoica,
  ``Kaluza-Klein relics from warped reheating,''
  Phys.\ Rev.\  D {\bf 77}, 123522 (2008)
  [arXiv:0710.1299 [hep-th]].

  J.~F.~Dufaux, L.~Kofman and M.~Peloso,
  ``Dangerous Angular KK/Glueball Relics in String Theory Cosmology,''
  Phys.\ Rev.\  D {\bf 78}, 023520 (2008)
  [arXiv:0802.2958 [hep-th]].

  A.~Buchel and L.~Kofman,
  ``'Black Universe' epoch in String Cosmology,''
  Phys.\ Rev.\  D {\bf 78}, 086002 (2008)
  [arXiv:0804.0584 [hep-th]].

\bibitem{post_inf}

  N.~Barnaby, A.~Berndsen, J.~M.~Cline and H.~Stoica,
  ``Overproduction of cosmic superstrings,''
  JHEP {\bf 0506}, 075 (2005)
  [arXiv:hep-th/0412095].

\bibitem{gw}

  J.~F.~Dufaux, A.~Bergman, G.~N.~Felder, L.~Kofman and J.~P.~Uzan,
  ``Theory and Numerics of Gravitational Waves from Preheating after
  Inflation,''
  Phys.\ Rev.\  D {\bf 76}, 123517 (2007)
  [arXiv:0707.0875 [astro-ph]].

  J.~F.~Dufaux, G.~N.~Felder, L.~Kofman and O.~Navros,
  ``Gravity Waves from Tachyonic Preheating after Hybrid Inflation,''
  JCAP {\bf 0903}, 001 (2009)
  [arXiv:0812.2917 [astro-ph]].

\bibitem{ng1}

  N.~Barnaby and J.~M.~Cline,
  ``Nongaussian and nonscale-invariant perturbations from tachyonic  preheating
  in hybrid inflation,''
  Phys.\ Rev.\  D {\bf 73}, 106012 (2006)
  [arXiv:astro-ph/0601481].

  N.~Barnaby and J.~M.~Cline,
  ``Nongaussianity from Tachyonic Preheating in Hybrid Inflation,''
  Phys.\ Rev.\  D {\bf 75}, 086004 (2007)
  [arXiv:astro-ph/0611750].

\bibitem{ng2}

  J.~R.~Bond, A.~V.~Frolov, Z.~Huang and L.~Kofman,
  ``Non-Gaussian Spikes from Chaotic Billiards in Inflation Preheating,''
  arXiv:0903.3407 [astro-ph.CO].

\bibitem{bigN}

  P.~Candelas, E.~Perevalov and G.~Rajesh,
  ``Toric geometry and enhanced gauge symmetry of F-theory/heterotic  vacua,''
  Nucl.\ Phys.\  B {\bf 507}, 445 (1997)
  [arXiv:hep-th/9704097].


\bibitem{Nflation}

  S.~Dimopoulos, S.~Kachru, J.~McGreevy and J.~G.~Wacker,
  ``N-flation,''
  JCAP {\bf 0808}, 003 (2008)
  [arXiv:hep-th/0507205].

\bibitem{matrix_spectrum}

  R.~Easther and L.~McAllister,
  ``Random matrices and the spectrum of N-flation,''
  JCAP {\bf 0605}, 018 (2006)
  [arXiv:hep-th/0512102].

\bibitem{Susskind}

  L.~Susskind,
  ``The anthropic landscape of string theory,''
  arXiv:hep-th/0302219.

\bibitem{multi_land}

  D.~Battefeld and T.~Battefeld,
  ``Multi-Field Inflation on the Landscape,''
  JCAP {\bf 0903}, 027 (2009)
  [arXiv:0812.0367 [hep-th]].

\bibitem{staggered}

  D.~Battefeld, T.~Battefeld and A.~C.~Davis,
  ``Staggered Multi-Field Inflation,''
  JCAP {\bf 0810}, 032 (2008)
  [arXiv:0806.1953 [hep-th]].

\bibitem{rand1}

  S.~H.~Tye, J.~Xu and Y.~Zhang,
  ``Multi-field Inflation with a Random Potential,''
  JCAP {\bf 0904}, 018 (2009)
  [arXiv:0812.1944 [hep-th]].

\bibitem{rand2}

  S.~H.~Tye and J.~Xu,
  ``A Meandering Inflaton,''
  Phys.\ Lett.\  B {\bf 683}, 326 (2010)
  [arXiv:0910.0849 [hep-th]].

\bibitem{stream}

  Y.~Wang,
  ``Multi-Stream Inflation: Bifurcations and Recombinations in the
  Multiverse,''
  arXiv:1001.0008 [hep-th].

\bibitem{assisted}

  A.~R.~Liddle, A.~Mazumdar and F.~E.~Schunck,
  ``Assisted inflation,''
  Phys.\ Rev.\  D {\bf 58}, 061301 (1998)
  [arXiv:astro-ph/9804177].

\bibitem{assisted2}

  P.~Kanti and K.~A.~Olive,
  ``On the realization of assisted inflation,''
  Phys.\ Rev.\  D {\bf 60}, 043502 (1999)
  [arXiv:hep-ph/9903524].

\bibitem{flat_scalar_pot}

  C.~P.~Burgess,
  ``Strings, branes and cosmology: What can we hope to learn?,''
  arXiv:hep-th/0606020.

\bibitem{lyth_bound}

  D.~H.~Lyth,
  ``What would we learn by detecting a gravitational wave signal in the  cosmic
  microwave background anisotropy?,''
  Phys.\ Rev.\ Lett.\  {\bf 78}, 1861 (1997)
  [arXiv:hep-ph/9606387].

  D.~Baumann and L.~McAllister,
  ``A Microscopic Limit on Gravitational Waves from D-brane Inflation,''
  Phys.\ Rev.\  D {\bf 75}, 123508 (2007)
  [arXiv:hep-th/0610285].



\bibitem{KSS}

  R.~Kallosh, N.~Sivanandam and M.~Soroush,
  ``Axion Inflation and Gravity Waves in String Theory,''
  Phys.\ Rev.\  D {\bf 77}, 043501 (2008)
  [arXiv:0710.3429 [hep-th]].

\bibitem{Nflation2}

  Y.~S.~Piao,
  ``On perturbation spectra of N-flation,''
  Phys.\ Rev.\  D {\bf 74}, 047302 (2006)
  [arXiv:gr-qc/0606034].

\bibitem{Nflation3}

  D.~Battefeld and T.~Battefeld,
  ``Non-Gaussianities in N-flation,''
  JCAP {\bf 0705}, 012 (2007)
  [arXiv:hep-th/0703012].


\bibitem{Nflation4}

  S.~A.~Kim and A.~R.~Liddle,
  ``Nflation: Multi-field inflationary dynamics and perturbations,''
  Phys.\ Rev.\  D {\bf 74}, 023513 (2006)
  [arXiv:astro-ph/0605604].

  S.~A.~Kim and A.~R.~Liddle,
  ``Nflation: Non-gaussianity in the horizon-crossing approximation,''
  Phys.\ Rev.\  D {\bf 74}, 063522 (2006)
  [arXiv:astro-ph/0608186].

  S.~A.~Kim and A.~R.~Liddle,
  ``Nflation: observable predictions from the random matrix mass spectrum,''
  Phys.\ Rev.\  D {\bf 76}, 063515 (2007)
  [arXiv:0707.1982 [astro-ph]].

\bibitem{multi_preheat}

  D.~Battefeld and S.~Kawai,
  ``Preheating after N-flation,''
  Phys.\ Rev.\  D {\bf 77}, 123507 (2008)
  [arXiv:0803.0321 [astro-ph]].

\bibitem{multi_preheat2}

  D.~Battefeld,
  ``Preheating after Multi-field Inflation,''
  Nucl.\ Phys.\ Proc.\ Suppl.\  {\bf 192-193}, 126 (2009)
  [arXiv:0809.3455 [astro-ph]].

  D.~Battefeld, T.~Battefeld and J.~T.~Giblin,
  ``On the Suppression of Parametric Resonance and the Viability of Tachyonic
  Preheating after Multi-Field Inflation,''
  Phys.\ Rev.\  D {\bf 79}, 123510 (2009)
  [arXiv:0904.2778 [astro-ph.CO]].

\bibitem{grimm}

  T.~W.~Grimm,
  ``Axion Inflation in Type II String Theory,''
  Phys.\ Rev.\  D {\bf 77}, 126007 (2008)
  [arXiv:0710.3883 [hep-th]].

\bibitem{nunes}

  J.~J.~Blanco-Pillado, D.~Buck, E.~J.~Copeland, M.~Gomez-Reino and N.~J.~Nunes,
  ``Kahler Moduli Inflation Revisited,''
  arXiv:0906.3711 [hep-th].

\bibitem{berglund}

  P.~Berglund and G.~Ren,
  ``Multi-Field Inflation from String Theory,''
  arXiv:0912.1397 [hep-th].

\bibitem{kklt}
  
  S.~Kachru, R.~Kallosh, A.~D.~Linde and S.~P.~Trivedi,
  ``De Sitter vacua in string theory,''
  Phys.\ Rev.\  D {\bf 68}, 046005 (2003)
  [arXiv:hep-th/0301240].

\bibitem{IIA}

  O.~DeWolfe, A.~Giryavets, S.~Kachru and W.~Taylor,
  ``Type IIA moduli stabilization,''
  JHEP {\bf 0507}, 066 (2005)
  [arXiv:hep-th/0505160].


\bibitem{green}

  D.~R.~Green,
  ``Reheating Closed String Inflation,''
  Phys.\ Rev.\  D {\bf 76}, 103504 (2007)
  [arXiv:0707.3832 [hep-th]].

\bibitem{vikman}

  G.~Dvali, I.~Sawicki and A.~Vikman,
  ``Dark Matter via Many Copies of the Standard Model,''
  JCAP {\bf 0908}, 009 (2009)
  [arXiv:0903.0660 [hep-th]].

\bibitem{race1}

  J.~J.~Blanco-Pillado {\it et al.},
  ``Racetrack inflation,''
  JHEP {\bf 0411}, 063 (2004)
  [arXiv:hep-th/0406230].

\bibitem{race2}

  J.~J.~Blanco-Pillado {\it et al.},
  ``Inflating in a better racetrack,''
  JHEP {\bf 0609}, 002 (2006)
  [arXiv:hep-th/0603129].

\bibitem{zhiqi}

  N.~Barnaby and Z.~Huang, unpublished note.

\bibitem{ref:Mathieu}

  N.~W.~McLachlan,
  ``Theory and Application of Mathieu Functions,''
  (Dover, New York, 1964).

\bibitem{restrap}

  H.~Broer, J.~Puig and C.~Simo,
  ``Resonance tongues and instability pockets in the quasi-periodic hill-schrodinger equation,''
  Comm. Math. Phys. {\bf 241}, no.2-3, pp. 467-503 (2003).

\bibitem{broer1}

  H.~Broer and C.~Simo,
  ``Hills equation with quasi-periodic forcing: resonance tongues, instability pockets and global phenomena,''
  Bol. Soc. Bras. Mat. {\bf 29}, pp. 253-293 (1998).

\bibitem{rrand1}

  R.~Zounes and R.~Rand,
  ``Transition curves for the quasi-periodic mathieu equation,''
  SIAM J. Appl. Math. {\bf 58}, 4, pp. 1094-1115 (1998).

\bibitem{rrand2}
  
  R.~Rand {\it et al.},
  ``A quasiperiodic mathieu equation,''
  Nonlinear Dynamics, The Richard Rand 50th Anniversary Volume, pp. 203-221,
  Series on Stability, Vibration and Control of Systems Series B: Vol. 2.

\bibitem{noise1}

  V.~Zanchin {\it et al.},
  ``Reheating in the presence of noise,''
  Phys. Rev. D {\bf 57}, 4651 (1998) 
  [arXiv:hep-ph/9709273].

\bibitem{noise2}

  V.~Zanchin {\it et al.},
  ``Reheating in the presence of inhomogeneous noise,''
  Phys. Rev. D {\bf 60}, 023505 (1999) 
  [arXiv:hep-ph/9901207].

\bibitem{cantor1}

  B.~Bassett and F.~Tamburini,
  ``Inflationary reheating in grand unified theories,''
  Phys. Rev. Lett. {\bf 81} (1998), 2630 
  [arXiv:hep-ph/9804453].

\bibitem{cantor2}

  B.~Bassett,
  ``Inflationary reheating classes via spectral methods,''
  Phys. Rev. D {\bf 58} (1998), 021303 
  [arXiv:hep-ph/9709443].

\bibitem{ref:preheatderiv}

  C.~Armendariz-Picon, M.~Trodden, and E.~West,
  ``Preheating in derivatively-coupled inflation models,''
  JCAP {\bf 0804},036 (2008) 
  [arXiv:0707.2177].

\bibitem{ref:steinhardt}
  
  I.~Zlatev, G.~Huey and P.~J.~Steinhardt,
  ``Parametric Resonance in an Expanding Universe,''
  Phys. Rev. D {\bf 57}, 2152 (1998) 
  [arXiv:astro-ph/9709006].

\bibitem{ref:tacres2}

  A.~A.~Abolhasani, H.~firouzjahi and M.~M.~Sheikh-Jabbari,
  ``Tachyonic Resonance Preheating in Expanding Universe,''
  Phys. Rev. D {\bf 81}, 043524 (2010)
  [arXiv:0912.1021 [hep-th]].

\bibitem{ref:kofman96}
  L.~A.~Kofman,
  ``The origin of matter in the universe: Reheating after inflation,''
  [arXiv:astro-ph/9605155].

\bibitem{ref:KLSpert}
  L.~A.~Kofman, A.~Linde, and A.~A.~Starobinsky,
  ``Reheating after inflation: 1. Perturbation Theory,''
  unpublished

\bibitem{eqn_state}

  D.~I.~Podolsky, G.~N.~Felder, L.~Kofman and M.~Peloso,
  ``Equation of state and beginning of thermalization after preheating,''
  Phys.\ Rev.\  D {\bf 73}, 023501 (2006)
  [arXiv:hep-ph/0507096].

\bibitem{yuki}

  Y.~Watanabe and E.~Komatsu,
  ``Gravitational inflaton decay and the hierarchy problem,''
  Phys.\ Rev.\  D {\bf 77}, 043514 (2008)
  [arXiv:0711.3442 [hep-th]].

\end{thebibliography}
\end{document}